\documentstyle[emulateapj]{article}

\makeatletter

\newenvironment{inlinefigure}{%
\def\@captype{figure}%
\noindent\begin{minipage}{0.999\linewidth}\begin{center}}
{\end{center}\end{minipage}\smallskip}
\makeatother

\begin{document}
\title{The faintest X-ray sources from $z=0-8$\altaffilmark{1,2,3}}
\author{
L.~L.~Cowie,$\!$\altaffilmark{4}
A.~J.~Barger,$\!$\altaffilmark{4,5,6}
G.~Hasinger$\!$\altaffilmark{4}
}

\lefthead{Cowie, Barger, \& Hasinger}

\altaffiltext{1}{Based in part on data obtained at the W. M. Keck
Observatory, which is operated as a scientific partnership among the/
the California Institute of Technology, the University of
California, and NASA and was made possible by the generous financial
support of the W. M. Keck Foundation.}
\altaffiltext{2}{Based in part on observations taken by the
CANDELS Multi-Cycle Treasury Program with the NASA/ESA HST, which is
operated by the Association of Universities for Research in Astronomy,
Inc., under NASA contract NAS5-26555.}
\altaffiltext{3}{Based in part on data obtained from the Multimission
Archive at the Space Telescope Science Institute (MAST).  STScI is
operated by the Association of Universities for Research in Astronomy,
Inc., under NASA contract NAS5-26555.  
}
\altaffiltext{4}{Institute for Astronomy, University of Hawaii,
2680 Woodlawn Drive, Honolulu, HI 96822.}
\altaffiltext{5}{Department of Astronomy, University of
Wisconsin-Madison, 475 North Charter Street, Madison, WI 53706.}
\altaffiltext{6}{Department of Physics and Astronomy,
University of Hawaii, 2505 Correa Road, Honolulu, HI 96822.}

\slugcomment{In press at The Astrophysical Journal}


\begin{abstract}
We use the new 4~Ms exposure of the CDF-S field obtained with 
the {\em Chandra\/} X-ray satellite to investigate the properties 
of the faintest X-ray sources over a wide range of redshifts.
We use an optimized averaging procedure to investigate the
weighted mean X-ray fluxes of optically selected
sources in the CDF-S over the redshift
range $z=0-8$ and down to $0.5-2$~keV fluxes as low
as $5\times10^{-19}$~erg~cm$^{-2}$~s$^{-1}$. None of the samples
of sources at high redshifts ($z>5$) show any significant
flux, and at $z=6.5$ we place an upper limit on the X-ray 
luminosity of $4\times10^{41}$~erg~s$^{-1}$ in the
rest-frame $3.75-15$~keV band for the sample of
Bouwens et al.\ (2006).  This is consistent with any X-ray 
production in the galaxies being solely due to star formation.
At lower redshifts we find significant weighted mean X-ray fluxes in 
many samples of sources over the redshift range $z=0-4$. 
We use these to argue
that (1) the relation between star formation and X-ray production
remains invariant over this redshift range, (2) X-ray sources
below the direct detection threshold in the CDF-S are primarily 
star-forming, and (3) there is full consistency between UV and X-ray
estimations of the star formation history.
\end{abstract}


\section{Introduction}
\label{intro}

We would like to be able to determine directly how much growth occurs 
in supermassive black holes in the early universe. Unfortunately, 
however, only the most luminous quasars at $z>6$ have ever been 
detected individually (e.g., Fan et al.\ 2006; Willott et al.\ 2010; 
Mortlock et al.\ 2011), and these sources are so massive that their 
evolutionary histories are far from typical. 

We would also like to know the relative contributions of the galaxy and 
active galactic nucleus (AGN) populations to the UV and X-ray ionizing 
radiation in the early universe.  Estimates from optical
data were that AGNs could not account for the required UV ionizing flux at
$z>3$ (e.g., Bolton et al.\ 2005; Meiksin 2005).  It is now recognized
that complete samples of AGNs---or
at least those which are not Compton thick---are most easily
obtained with X-ray observations.  Direct searches for 
high-redshift ($z=3.5-6.5$) AGNs carried out using combined deep 
{\em Chandra\/} X-ray samples and optical imaging data
(e.g., Barger et al.\ 2003; Fontanot et al.\ 2007) also found
that the ionization from AGNs was insufficient to maintain the
observed ionization of the intergalactic medium at high redshifts.
This result was conclusively established at $z>3$ by Cowie et al.\ (2009) 
by measuring the contribution of AGNs to the ionizing flux 
as a function of redshift using X-ray, optical, and UV observations.

However, even if the space-density of high-redshift, unobscured AGNs is
not sufficiently high to reionize the universe, it does not rule out
the possibility that substantial accretion onto supermassive 
black holes is occurring behind veils of obscuring material.  
The only way to determine this is by probing the faintest X-ray fluxes.
Thus, the continued deepening of the deepest X-ray images of the sky
is essential for studying this important issue.
The latest of these deepenings is the now nearly 4~Ms of exposure on the 
{\em Chandra\/} Deep Field-South
(CDF-S; Giaconni et al.\ 2002; Alexander et al.\ 2003, hereafter A03; 
Luo et al.\ 2008; Xue et al.\ 2011, hereafter X11). 
The 4~Ms exposure is deep enough to detect all sources above
a luminosity of $\sim5\times10^{42}$~erg~s$^{-1}$
in the rest-frame $3.5-14$~keV band at $z=6$.  Thus,
almost all luminous AGNs, even at very high
redshifts, will be included in the catalog
of directly detected sources. However, only one source
with a redshift greater than $z=5$ has been spectroscopically
identified (Barger et al.\ 2003) in the {\em Chandra\/} Deep Field-North 
(CDF-N; Brandt et al.\ 2001; A03)  and CDF-S fields.

Fainter sources are likely to be dominated by star formation,
low-luminosity AGNs (LLAGNs), or highly
obscured AGNs. Such sources should have
strong breaks in the rest-frame UV 
owing to the Ly$\alpha$ forest of the intergalactic
gas or to the intrinsic Lyman continuum break and should be
included in optical or near-infrared
dropout samples identified from the break properties.
We can attempt to measure the properties of these fainter
sources through averaging (sometimes referred to as ``stacking'')
analyses of optically or near-infrared 
selected galaxy samples (e.g., Brandt et al.\ 2001;
Nandra et al.\ 2002; Reddy \& Steidel 2004; Lehmer et al.\ 2005;
Worsley et al.\ 2006; Hickox \& Markevitch 2006, 2007; Treister et al.\ 2011). 
This is of particular interest for the highest redshift samples, where 
we may be probing supermassive black holes at times very close to their 
formation.  However, it also allows us to examine the X-ray properties 
of star-forming galaxies over a wide range in redshifts
(out to beyond $z=4$) and to test whether the properties
of these galaxies change with time.

Treister et al.\ (2011; hereafter T11) used the CDF-N and CDF-S 
data to perform sensitive X-ray stacking analyses 
on $z>6$ galaxy candidates 
that could not be detected individually to look for detectable 
X-ray signal from the population as a whole.
The high-redshift samples they used ($z\sim6$ from Bouwens et al.\ 2006,
hereafter Bo06, and $z\sim 7$ and $z\sim 8$ from Bouwens et al.\ 2011,
hereafter Bo11) were selected 
with the drop-out technique using {\em Hubble Space Telescope (HST)\/} 
ACS and WFC3 data.  For the $z\sim6$ Bo06 sample,
T11 found significant detections ($\ge5\sigma$) in both 
the observed soft ($0.5-2$~keV) and the observed hard ($2-8$~keV) bands. 
The stacked hard-to-soft X-ray flux ratio that they found was a relatively 
high factor of $\sim9$, which they said could only be explained
by very high levels of obscuration.  Moreover, since the ratio was a ratio
of the stacks, they inferred that there must be very few sources with 
significantly lower levels of obscuration.  This led them to the
conclusion that black holes grow significantly at early times but are 
heavily buried in gas and dust.

\begin{figure*}
\centerline{\includegraphics[width=4.0in,angle=90]{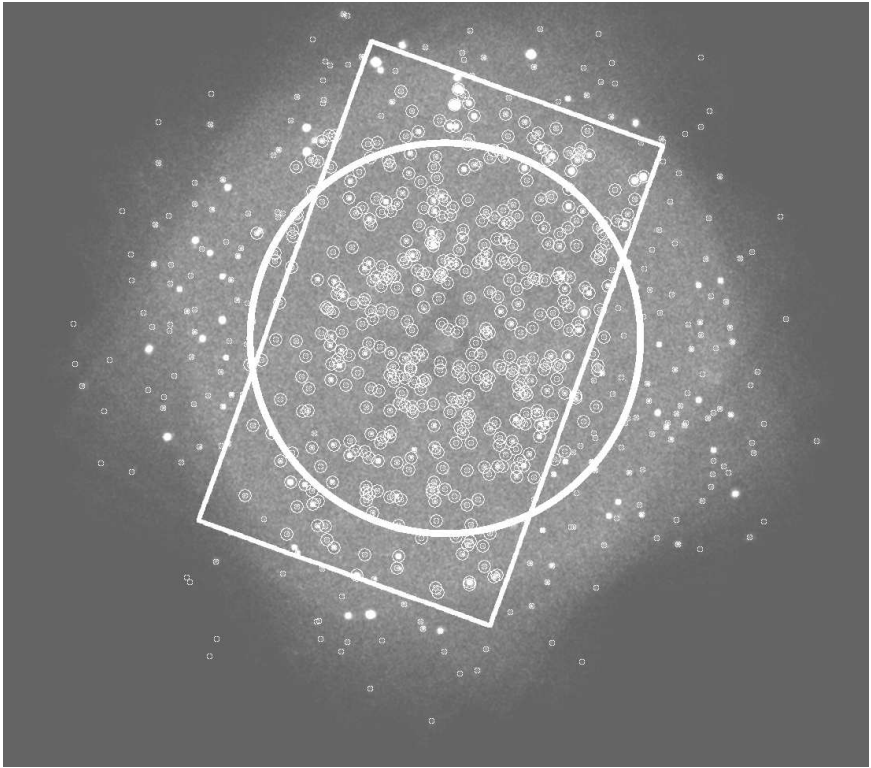}}
\caption{
CDF-S 4~Ms full-band image overlaid with a circle showing an
off-axis radius of $6'$ and a rectangle showing the {\em HST\/} ACS
uniformly covered region of the GOODS-S (Giavalisco et al.\ 2004).
The small circles show the positions of the X-ray sources in the 
X11 CDF-S catalog, and the larger circles show 
the X-ray sources with NIR coverage from the ERS (Windhorst et al.\ 2011) 
and CANDELS (Grogin et al.\ 2011; Koekemoer et al.\ 2011)
observations using the F125W and F160W filters on {\em HST\/} WFC3.
We refer to the area within both the $6'$ radius circle and the uniformly 
covered region of the GOODS-S as our core region. 
\label{image}
}
\end{figure*}

These results are rather surprising. 
Galaxies selected by the drop-out technique are generally classic 
starburst galaxies; i.e., they are small, blue, and compact with high 
surface brightnesses (e.g., Stanway et al.\ 2004).  
Meanwhile, the masses of local supermassive 
black holes are generally correlated with the bulge
luminosities of the host galaxies 
(e.g., Ferrarese \& Merritt 2000; Kormendy \& Gebhardt 2001; 
Gebhardt et al.\ 2000; Tremaine et al.\ 2002).
At $z\sim3$, this would be akin to finding AGNs
in Lyman Break Galaxies (LBGs), which is not very common 
(e.g., Steidel et al.\ 2004).
Moreover, if there were a lot of gas and dust present to obscure 
the central black hole (T11 argue that the sources must be
nearly Compton-thick along most directions), then one might 
expect the dust to also affect the properties of the host galaxy,  
unless it were somehow confined to the region around the AGN itself.

Indeed, Fiore et al.\ (2011) conducted an independent, non-optimized 
stacking analysis using the Bo06 sample in the CDF-S and did not 
confirm the T11 results.  More recently, Willott (2011) provided
an independent critique of T11 that parallels some of the
results of the present paper. 

Here we argue that the T11 result is a 
consequence of an incorrect background subtraction procedure. 
In our analysis we find a $2\sigma$ upper limit on the X-ray 
luminosity at $z=6.5$ of $4\times10^{41}$~erg~s$^{-1}$, which is
consistent with any X-ray production in the galaxies being solely 
due to star formation.

At lower redshifts we find significant weighted mean X-ray fluxes 
in many samples 
of sources over the redshift range $z=0-4$. We use these to argue
that (1) the relation between star formation and X-ray production
remains invariant over this redshift range, (2) X-ray sources
below the direct detection threshold are primarily star-forming,
and (3) there is full consistency between UV and X-ray
estimations of the star formation history.

We use a standard $H_0=70$~km~s$^{-1}$~Mpc$^{-1}$, 
$\Omega=0.3$, $\Omega=0.7$ cosmology throughout. All magnitudes are 
in the AB magnitude system.

\section{X-ray Data}
\label{xray}

X11 presented the X-ray images from 
the 3.872~Ms total exposure of the CDF-S, which covers an area of 
464.5~arcmin$^2$.  They provided a main {\em Chandra\/}
source catalog containing 740 X-ray sources detected with WAVDETECT
at a false-positive probability threshold of $10^{-5}$ in at least one
of three X-ray bands:  full ($0.5-8$~keV), soft ($0.5-2$~keV),
and hard ($2-8$~keV).  The catalog also satisfies a binomial-probability
source-selection criterion of $P<0.004$, which means that the probability
that the sources not are real is less than 0.004. 
The on-axis flux limits are
$\approx3.2\times 10^{-17}$, $9.1\times 10^{-18}$, and 
$5.5\times 10^{-17}$~erg~cm$^{-2}$~s$^{-1}$ for the full, soft, and
hard bands, respectively. 

The CDF-S is now almost twice
as deep as the CDF-N field, which is the only other
extremely deep {\em Chandra\/} field. For most of our analysis
we will therefore concentrate on the CDF-S alone, but in
some cases we will also include similarly selected samples
from the CDF-N, where these can provide an increase
in the sensitivity. For the CDF-N we use the images
and corresponding source catalog from A03, 
which contains 503 sources selected
in the same three X-ray bands described above.

X11 shifted the optical positions in deep $R$-band imaging data---which 
were already matched to the {\em HST\/} ACS Great Observatories
Origins Deep Survey-South (GOODS-S; Giavalisco et al.\ 2004)
images---by $0\farcs175$ in right ascension and $-0\farcs284$ in 
declination in order to match them to the radio positions of sources 
in the Very Large Array (VLA) 1.4~GHz radio catalog ($5\sigma$) of 
Miller et al.\ (2008) and N.~A.~Miller et al.\ (2011), in preparation.
They then matched the X-ray centroid positions to the optical sources 
detected in the $R$-band image to put them on a common astrometric frame.  
In this paper, we moved X11's source positions back to the 
GOODS-S ACS frame by shifting them
by $-0\farcs175$ in right ascension and $0\farcs284$ in declination.
This allows us to work with the astrometric positions of the optical 
sources and spectral data in the GOODS-S samples.

In Figure~\ref{image} we show the X-ray image of the CDF-S. 
With small circles we show the positions
of all of the X-ray sources from X11.  With larger circles we
show the positions of the X-ray sources
that also have deep near-infrared (NIR) coverage from {\em HST\/} 
WFC3 imaging (Early Release Science or ERS, Windhorst et al.\ 2011;
Cosmic Assembly Near-Infrared Deep Extragalactic Legacy Survey 
or CANDELS, Grogin et al.\ 2011 and Koekemoer et al.\ 2011).  
Given the X-ray positional uncertainties, we identified an
X-ray source with a NIR counterpart if an F160W band 
counterpart brighter than F160W$\sim25$ ($5\sigma$) was within
$2''$ of the X-ray position.  If more than one such NIR
counterpart was within the search radius, then we identified
the X-ray source with the nearest NIR neighbor.
As pointed out in X11 (see also A.~J.~Barger et al.\ 2011, in preparation),
nearly all of the X-ray sources have NIR counterparts.
Finally, we mark with a rectangle the part of the GOODS-S region that 
was uniformly covered with ACS. The X-ray sensitivity degrades rapidly
at large off-axis radii as the point spread function (PSF)
of the X-ray image increases, so we restrict our analysis
to sources lying within an off-axis radius of $6'$ (very large circle).
The overlap of the $6'$ region with the uniformly covered GOODS-S 
region is 101~arcmin$^2$. 
This region contains 360 of the 740 X-ray sources in X11 
and 19,294 of the 33,955 optically selected sources 
given in the v2 catalog of the ACS observations of the GOODS-S.
In this paper we shall refer to this overlap region in both fields 
as our core region.

\section{Target Samples Over the $z=0-8$ Redshift Range}
\label{targets}

We measured the weighted mean X-ray counts~s$^{-1}$ in three 
types of selected galaxy samples. First,
we used galaxies with secure spectroscopic redshifts.
This is a relatively small sample, and the sources
are bright in the rest-frame optical. Thus, if
more optically-luminous galaxies are X-ray brighter
(e.g., Lehmer et al.\ 2005), then this sample may 
be expected to produce higher weighted mean X-ray counts~s$^{-1}$. 
Second, we used a photometric redshift
sample to probe fainter galaxies. Finally, we used dropout
selected samples to provide as complete a selection
as possible in a given redshift range.

\subsection{Secure Spectroscopic Redshift Sample}
\label{spectroscopic}

For the GOODS-S, we started with the 
European Southern Observatory (ESO) master redshift catalog 
compiled by A.~Rettura in 2004 and subsequently updated by 
Popesso et al.\ (2009) and Balestra et al.\ (2010).
This catalog includes redshifts from
Cristiani et al.\ (2000),
Croom et al.\ (2001),
Bunker et al.\ (2003),
Dickinson et al.\ (2004),
Stanway et al.\ (2004),
Strolger et al.\ (2004),
Szokoly et al.\ (2004),
van der Wel et al.\ (2004),
Doherty et al.\ (2005),
Le F{\`e}vre et al.\ (2005),
Mignoli et al.\ (2005), 
Ravikumar et al.\ (2007), and
Vanzella et al.\ (2008).
We added to this catalog
144 new redshifts for previously unidentified
sources that we measured with DEIMOS in the Fall of 2010, as well
some additional redshifts obtained by S. Koposov et al.\ (unpublished) 
that were presented in Taylor et al.\ (2009).
In our compilation
we only included sources with secure redshifts. In the case of the
VIMOS-VLT Deep Survey or VVDS (Le F{\`e}vre et al.\ 2005), we only 
included sources classified as 100\% secure. 
We included a small number of redshifts from papers 
that did not give a quality flag, but in these cases the spectral 
identifications appeared robust based on the material presented
in the papers.

In our final compilation, we have secure redshifts for 1792 
galaxies and 158 stars in the full GOODS-S field. 
In Figure~\ref{zplot}(a) we plot the spectroscopic redshifts 
versus the F850LP magnitudes for the final compilation
(black small squares). 
1158 of the galaxies with secure redshifts lie in the core
region. Using insecure redshifts would add
a comparable number of additional redshifts, but many of these
are clearly problematic based on our inspections of the spectra
or on our comparisons of the redshifts given in the different 
catalogs, so we prefer to work only with the secure redshifts.

For the GOODS-N, we used the catalogs of Barger et al.\ (2008)
updated with as yet unpublished DEIMOS redshifts obtained by
us and some additional redshifts from
Cooper et al.\ (2011). The GOODS-N sample has secure redshifts 
for 2987 galaxies and 199 stars. The slightly larger sample in
the GOODS-N as compared to the GOODS-S compensates for the shallower 
exposure in the CDF-N, and we use both samples in computing the 
weighted mean fluxes.

\begin{inlinefigure}
\includegraphics[width=2.8in,angle=90]{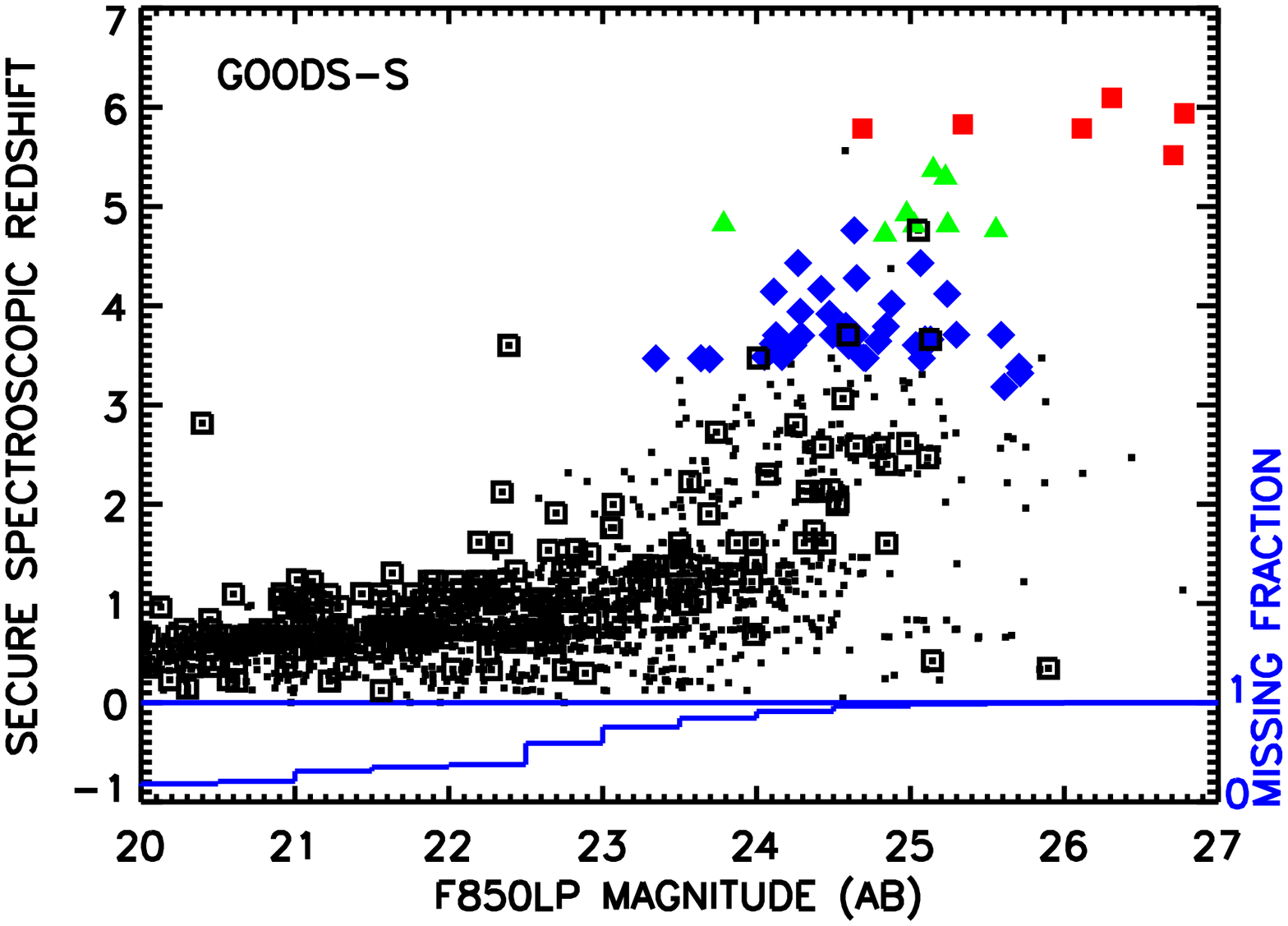}
\includegraphics[width=2.8in,angle=90]{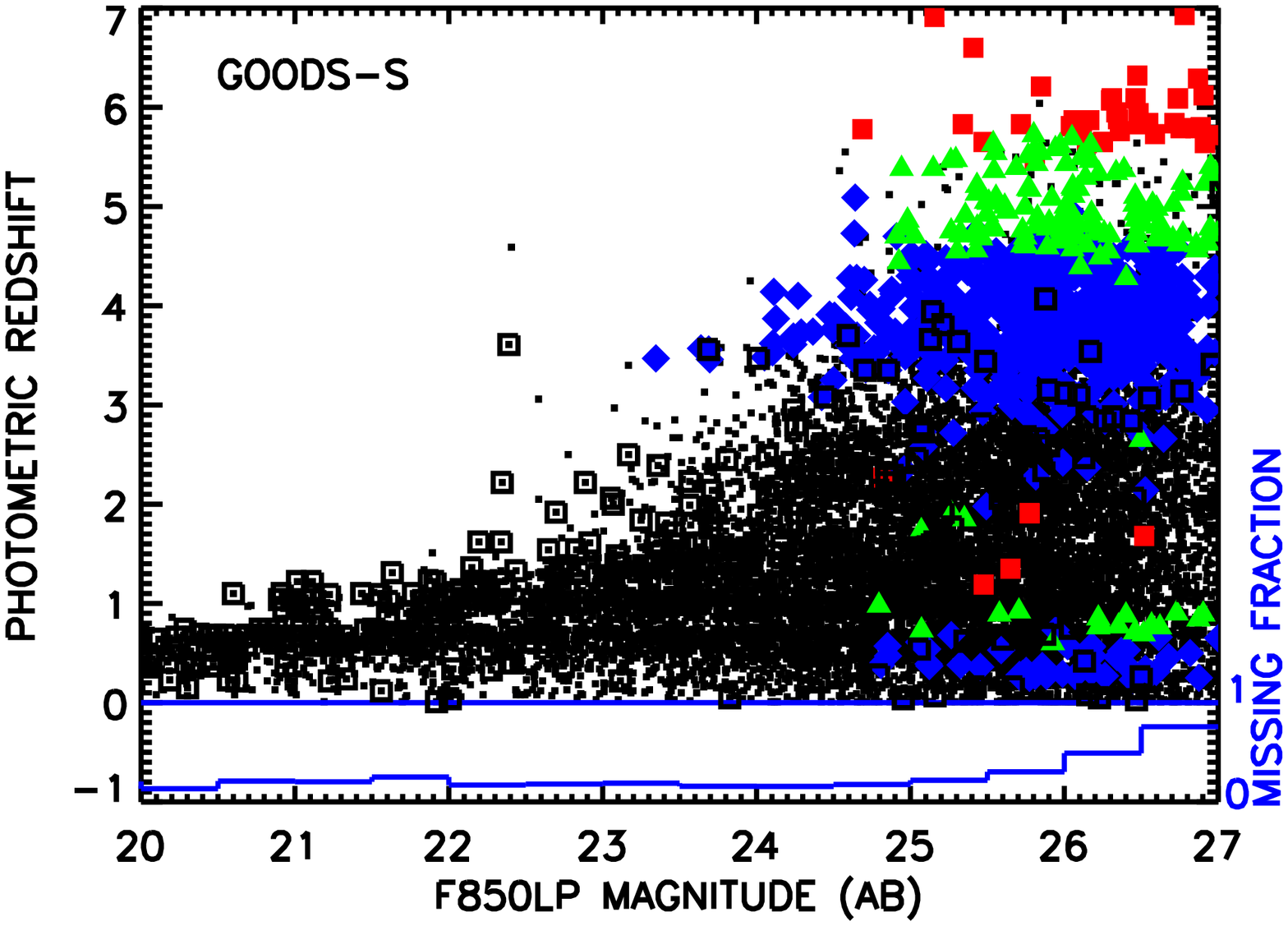}
\caption{
(a) Secure spectroscopic redshift vs. F850LP magnitude for all the 
sources with secure spectroscopic redshifts in the full GOODS-S
(black small squares; see text for references). 
X-ray sources are marked with a black enclosing square.
The lower blue histogram (right-hand vertical scale) shows the fraction 
of sources that have not been identified as a function of magnitude. 
(b) Same for the sources with photometric redshifts
from Grazian et al.\ (2006).
The colored symbols show the $b$ (blue), $v$
(green), and $i$ (red) dropouts from Beck06 (plus Bo06
for the $i$ dropout sample) that have
(a) spectroscopic redshifts or (b) photometric redshifts. 
The more X-ray luminous broad-line AGNs are not
selected by the dropout techniques.
\label{zplot}
}
\end{inlinefigure}

\subsection{Photometric Redshift Sample}
\label{photz}

For our photometric redshift sample, we restricted to the deeper 
CDF-S data and used the GOODS-MUSIC catalog of Grazian et al.\ (2006).
(Dahlen et al.\ 2010 report on an alternate photometric
redshift computation based on updated data, 
but the actual catalog has yet to be released.)
The Grazian et al.\ catalog gives photometric redshifts 
for 13,820 sources in the full GOODS-S field. They
reproduce the secure spectroscopic redshifts quite well, 
though there are a fair number of outliers: 
about 4\% of sources lie
at $(z-z_p)/(1+z) > 0.2$. Much of this seems to be associated
with the $z=2-3$ redshift range. At higher redshifts there
are few problems, presumably because the identifications
are more secure due to the breaks.  In Figure~\ref{zplot}(b)
we show the photometric redshifts versus the F850LP magnitudes 
for the Grazian et al.\ (2006) sample (black small squares). 
This sample contains 10,071 galaxies in our core region.

\subsection{Dropout Samples}
\label{drops}

We used Beckwith et al.\ (2006, hereafter Beck06)'s list of $b$
(1335), $v$ (328), and $i$ (105) dropouts in the GOODS-S and 
Hubble Ultra Deep Field (HUDF). Bo06 compiled an alternative list 
of 522 $i$ dropouts in these fields plus the GOODS-N.
The Beck06 $i$ dropout sample overlaps 
with the Bo06 sample but contains some sources that are 
not in Bo06 and vice versa. We will consider
both of these samples separately in our subsequent analysis.

We show the sources in these dropout selected samples
in the redshift-magnitude plots of Figure~\ref{zplot}
(colored symbols).
All of the spectroscopically identified sources fall
at the redshifts expected from the dropout selection. 
There is a small amount
of inconsistency between the dropouts and the photometric
redshifts, as can be seen in Figure~\ref{zplot}(b), where
a number of the dropouts lie at low photometric
redshifts. The inconsistent fraction is 68 out of
631 objects with photometric redshifts in the $b$ dropout
sample, 33 out of 161 in the $v$ dropout sample, and 6 out
of 39 in the $i$ dropout sample.
The median photometric redshifts (which we will use for
later plots) are $z=3.74$ for the $b$ dropouts, 
$z=4.78$ for the $v$ dropouts, and $=5.84$ for the $i$ dropouts 
(where the latter contains both the Beck06 and Bo06 samples).

Higher redshift ($z\sim7-8$) sources fall out of the GOODS-S 
F850LP selected sample.  Here we use the sample of Bo11.
Although McLure et al.\ (2011)
criticized some of the Bo11 sources (which they consider
to be lower redshift interlopers) and gave their own
robust list, we use the Bo11 samples here because they are deeper.
However, we note that we do not detect any significant mean X-ray flux
in either sample and that the error limits are similar regardless
of the choice of sample.

\section{Optimized Averaging Procedure}
\label{stack}

A fully optimally weighted determination of the
mean X-ray flux in a sample from the {\em Chandra\/} data would involve 
using variable elliptical apertures that are dynamically adjusted 
to match the local PSFs and background.
However, with optimal weightings, most of the signal---even for a 
set of sources that is uniformly distributed over the full {\em Chandra\/}
field---arises at relatively small off-axis radii
(less than $6'$), where we may include most of the counts within
the PSF with a moderately sized circular aperture
(e.g., Allen et al.\ 2004). Even at these small off-axis
angles the optimal weighting of the {\em Chandra\/} data involves 
several issues, most particularly which aperture to choose and 
how far out in off-axis angle to include sources 
(e.g., Lehmer et al.\ 2005; Hickox \& Markevitch 2006, 2007; T11).

For faint sources, the noise level is dominated by the background. 
We may write the S/N in an aperture of radius $r$ as proportional
to $f(r,\theta)/(b^{0.5}\times r)$, where $f(r,\theta)$ is
the fraction of the counts from the source 
contained within the aperture at off-axis radius $\theta$
(often referred to as the enclosed energy fraction),
and $b$ is the background per unit solid angle (see also T11).
We show this quantity for the $2-8$~keV band in Figure~\ref{sn_figure}
as a function of $\theta$ (black solid: $0'$ or on-axis;
red: $1.5'$; blue: $3'$; green: $5'$; black dashed: $7'$)
and $r$, assuming $b$ is roughly constant
within the region. As T11 point out, the optimal $r$ to maximize
the S/N increases with increasing $\theta$, rising from
$r=0.4''$ at $\theta=0'$ (black solid) to 
approximately $r=5''$ at $\theta = 7'$ (black dashed).
However, as can also be seen from Figure~\ref{sn_figure},
the dependence on $r$ at larger $\theta$ is very soft, so
choosing a smaller aperture radius has little effect on the S/N.
Moreover, using a smaller aperture radius has the advantage of 
minimizing contamination by neighboring sources. 

A very small aperture is optimal close to on-axis, but 
the use of such a small aperture places stress on the astrometry
(see T11). We have therefore chosen to use a constant $r=0.75''$
aperture within an off-axis angle of $\theta=3'$ 
and a constant $r=1.25''$ aperture at larger $\theta$.
This roughly brackets the optimal radius for $\theta=3'$
(see blue curve in Figure~\ref{sn_figure} for the $2-8$~keV band), 
namely $r=0.9''$ in the $0.5-2$~keV band and $r=1.2''$ 
in the $2-8$~keV band. The noise in typical samples
computed with this procedure only differs from that computed
with a fully optimized aperture as a function of radius by a 
few percent. 

We include sources out to $\theta=6'$, 
which, in conjunction with the samples lying within
the well covered areas of the GOODS
fields, is our definition of the core region.
We stress that the results are quite insensitive to the 
choice of the maximum $\theta$, 
because the high off-axis sources have low S/N
in an optimally weighted mean.
Decreasing the off-axis angle at which we include sources
to $\theta=5'$ ($4'$) would raise the noise in typical 
samples by 5\% (15\%).

\begin{inlinefigure}
\hskip 2.5cm
\includegraphics[width=2.8in,angle=90]{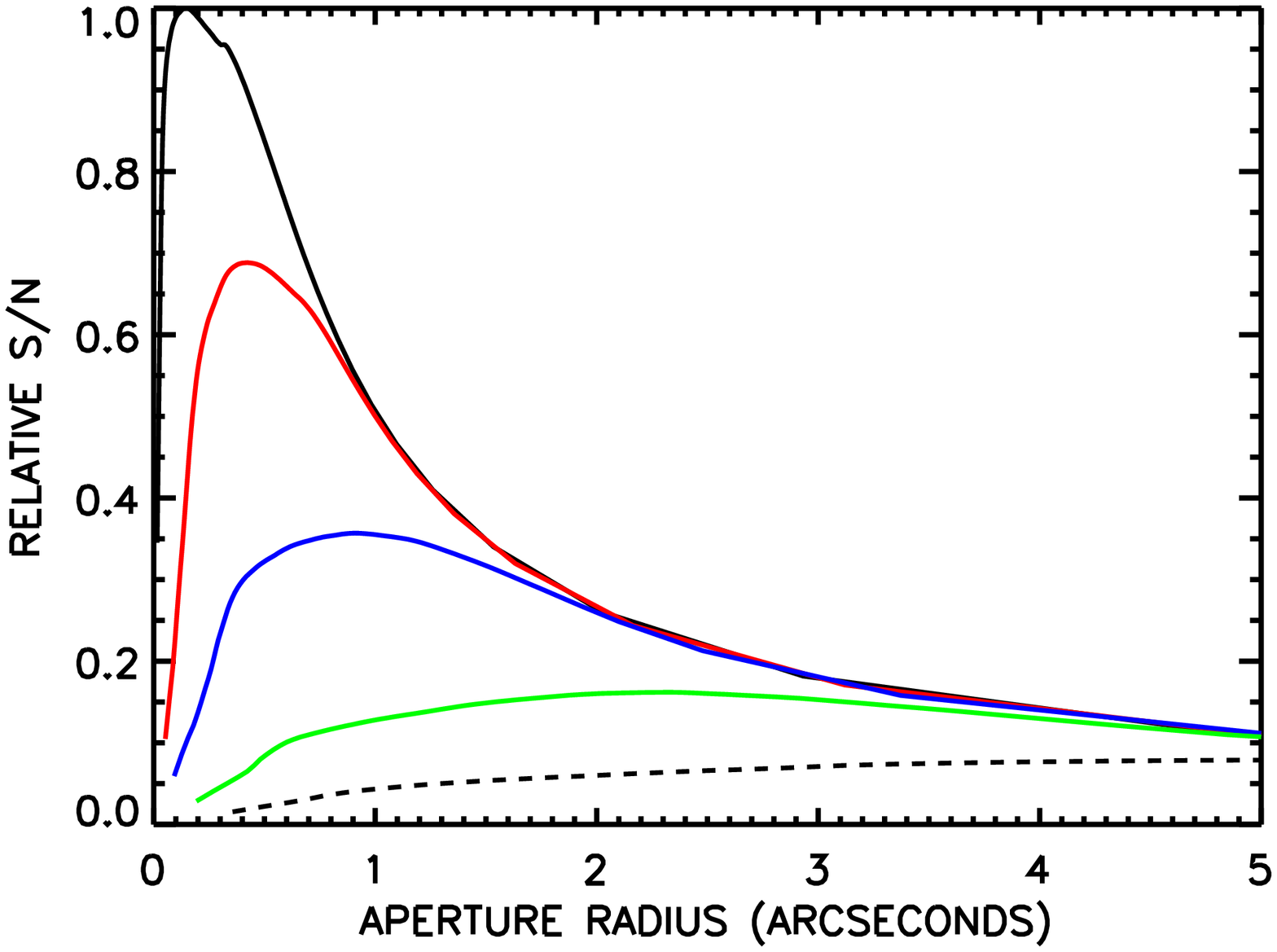}
\caption{Relative S/N in the $2-8$~keV band vs.
aperture radius shown for five values of $\theta$: $0'$ 
or on-axis (black solid curve), $1.5'$ (red), $3'$ (blue), 
$5'$ (green), and $7'$ (black dashed).
The value is normalized to the peak on-axis value.
\label{sn_figure}
}
\end{inlinefigure}

With the aperture specified, we can now compute the X-ray
counts~s$^{-1}$ as $C = (S-B)/(f(r,\theta) \times t$),
where $S$ is the number of counts in the aperture,
$B(=\pi r^2 b)$ is the number of background counts expected
in the same aperture, 
and $t$ is the effective exposure time at the 
position of this aperture. $C$ may be negative or positive.
We take the error on this quantity to be
$E = B^{0.5}/(f(r,\theta) \times t)$, since, in
general, the number of background counts in the aperture, $B$, 
is high enough to justify the Gaussian approximation.

For a given sample we form the optimally weighted mean X-ray
counts~s$^{-1}$, $L$, and the corresponding error, $E_L$. 
To do this, we first exclude all sources with a known
X-ray source within $6''$ or a bright ($0.5-8$~keV
counts above 1000) X-ray source within $13''$ in the existing 
catalogs (X11 for the CDF-S and A03
for the CDF-N) to remove any contamination
from these sources.  These two values ($6''$ and $13''$) 
correspond to the 90\% and 95\% enclosed light radii,
respectively, at an energy of 5~keV and an off-axis radius
of $6'$. This ensures that the contribution
from the extended wings of the known sources is
less than about a count in any aperture, even for the hard 
band.  When averaged over the ensembles, this level
of contamination is negligible. 
Our contamination radii are smaller than 
the values adopted by Hickox \& Markevitch (2006, 2007) or 
by T11, but tests with random samples show that there is
little sensitivity to increasing them,
and choosing the smallest possible contamination radii
minimizes the number of sources excluded by this
step. The adopted procedure reduces the
core area from 101~arcmin$^2$ to 86~arcmin$^2$.
We form the optimally weighted mean X-ray counts~s$^{-1}$, 
$L = \Sigma (C/E^2)/\Sigma(1/E^2)$, by
summing over the remaining sources in the sample.
The corresponding error is $E_L = (1/\Sigma(1/E^2))^{0.5}$.

The final issue is how to compute the background counts per
unit solid angle, $b$.
There are two possible approaches to this. One can
compute a local background for each source using either
a surrounding annulus (T11) or a wavelet approach
(Hickox \& Markevitch 2006, 2007), or one can compute 
the average background by randomizing the positions
of the sources and measuring the values obtained
in the blank-field apertures generated (e.g., Lehmer et al.\ 2005).
Here we use both procedures: We use annular
measurements as our primary background determination, and
then we use randomized position measurements to test that
the procedure is not introducing offsets.

Before we continue, we note that the background we are measuring
is composed of several elements: an approximately
uniform instrumental background, a truly diffuse
component containing contributions
from hot interstellar or intergalactic gas, and contributions
from individual sources below the direct detection threshold
of the deep CDF images. While a local estimate of the
background contributions from the instrument and the diffuse background
should provide a good average measure, such estimates could be
more problematic for the unresolved sources, which may be clustered.
Clustering could, in principle, result in an underestimate in determining
the background subtraction, since the annulus in which the background
is measured lies further from the source than the aperture in which
the signal is measured. However, in practice, the surface density of 
X-ray sources at a given redshift is small, and the average 
contribution of these sources to the background
is also small. (Contributions from sources at other redshifts may
be viewed as random with respect to the sources whose signal is being
measured.) 

We may use the results of Section~\ref{photzsamp} to 
estimate that the total contribution of all the sources in our 
photometric redshift sample is approximately 2\% of the observed 
background in the CDF-S.
In smaller redshift intervals, these contributions are correspondingly
less and drop rapidly with increasing redshift
(see Section~\ref{discuss}). Thus, the redshift interval from $z=1-1.1$
produces about 0.001 of the background, and that at $z=3-3.1$ produces
less than  0.0003 of the background. These contributions are too small for 
clustering in the galaxy population to perturb the background in any 
significant way.

In determining $b$, we use an annulus of
radius $8''$ to $22''$.  This choice of inner boundary
avoids any contribution from the target, and this choice of 
outer boundary provides a large enough number of pixels. 
In order to eliminate any contribution to $b$ from bright 
neighboring sources, we exclude $6''$ regions around known 
sources ($13''$ regions around very bright sources), and we 
clip pixels with high counts. Considerable caution must be taken
in performing the clip. Because the average number of counts 
in each pixel is small, the counts distribution is 
Poissonian. Too severe of a clip can introduce
a downward bias in the estimation of the 
average background in the annulus, which would
result in a spurious signal when that background is
subtracted from the counts in the aperture.
This appears to be the cause of the $z\sim6$ weighted 
mean signal found by T11.

\begin{inlinefigure}
\hskip 2.5cm
\includegraphics[width=2.8in,angle=90]{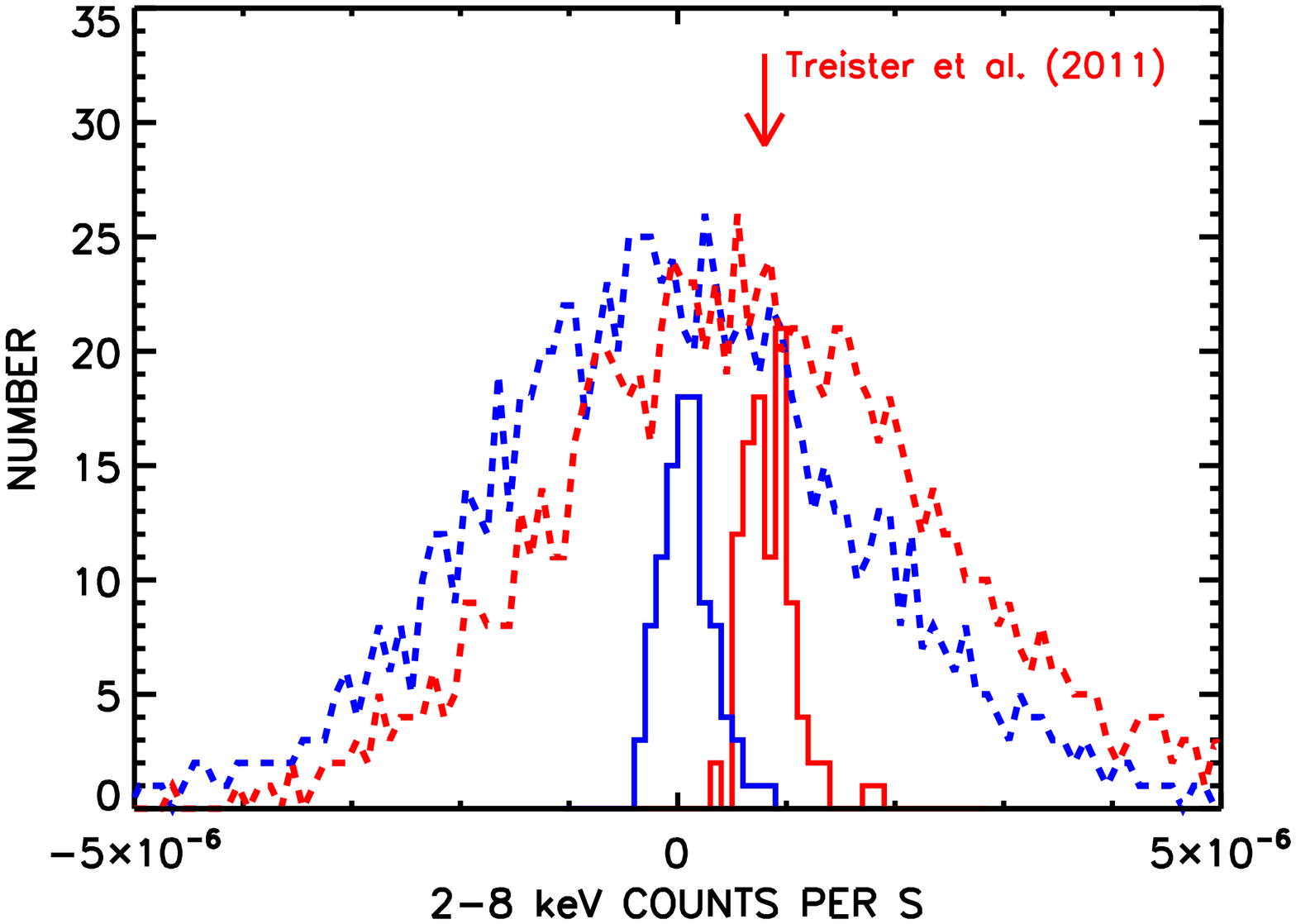}
\caption{Two distributions of $2-8$~keV counts~s$^{-1}$, $C$,
from 100 realizations of randomly located apertures in the
GOODS-S core region for a sample with a size equal to that of 
the Bo06 sample of $z\sim6$ galaxies.
The red dashed curve shows the results 
when the background pixels are clipped above 2 counts 
(this corresponds to the T11 procedure), 
and the blue dashed curve shows the results when the background 
pixels are instead only clipped above 4 counts (present procedure).
The solid line histograms show the distributions of 
the weighted mean $2-8$~keV counts~s$^{-1}$, $L$.
The more severe clipping procedure of T11 (red) introduces an offset 
in the distribution that is almost identical to the value measured in 
the actual Bo06 sample by T11 (downward pointing arrow).
\label{counts_hist}
}
\end{inlinefigure}

To test both the present procedure and the T11 procedure, 
we ran 100 realizations 
of randomly located apertures in the GOODS-S core region
for a sample with a size equal to that of the Bo06 sample
of $z\sim6$ galaxies.  
In Figure~\ref{counts_hist} we show the two distributions of 
the $2-8$~keV counts~s$^{-1}$, $C$, that we measured.
In one case we clipped the background pixels above 2 counts
(red dashed curve; this corresponds to the T11 procedure), 
while in the other case we only clipped the background pixels
above 4 counts (blue dashed curve; present procedure). 
The more severe clipping (red) produces a systematic 
offset in the measured counts, while the less severe clipping 
(blue) does not. The 4 counts clip used in the present paper
is the most stringent clip possible that does not produce
a significant offset. However, we note that the results are
not sensitive to choosing higher values for the clipping.

We can see this offset even more clearly when we plot the 
two distributions of weighted mean $2-8$~keV counts~s$^{-1}$, $L$. 
(Red solid for the T11 procedure; blue solid for the 
present procedure.) The offset of the red solid histogram
($8.1\times10^{-7}\pm 2.2\times10^{-7}$ counts~s$^{-1}$) 
is almost identical to the weighted mean signal found by T11 
for the Bo06 sample (downward pointing arrow).
In contrast, the offset of the blue solid histogram
($5\times10^{-8}\pm 2.2\times10^{-7}$ counts~s$^{-1}$) is 
consistent with the zero weighted mean signal expected for 
random realizations if the background subtraction is correct. 

The error in the simulations measured from the dispersion
in the realizations is about 30\%
higher than the formal statistical error of 
$1.6\times10^{-7}$ counts~s$^{-1}$,
suggesting that systematic effects may be present. 
Thus, we allow for this in assessing the significance of 
the detections.

\section{Results}
\label{results}

\subsection{Secure Spectroscopic Redshift Sample}
\label{specsamp}

In Figure~\ref{xray_redshift} we show the weighted mean
counts~s$^{-1}$ in both the $0.5-2$~keV (red squares) and 
$2-8$~keV (blue diamonds) bands versus redshift
for the core samples with secure spectroscopic redshifts.
In (a) and (b) we show the results, respectively, for the 
GOODS-S and GOODS-N fields, and in (c) we show the
results for the two fields combined.
The individual fields show a broadly
similar pattern. As has been found in previous work 
(e.g., Lehmer et al.\ 2005), there is a strongly detected 
signal in the $0.5-2$~keV band out to $z=4$ 
(see also Bomans, Zinn and Blex\ 2011).
Beyond this redshift, even in the combined fields, the signal 
falls below the $3\sigma$ threshold, even when only
the formal statistical error is used.  

The weighted mean $0.5-2$~keV counts~s$^{-1}$ in the combined fields
are $5.2\pm1.0\times10^{-7}$ counts~s$^{-1}$ 
for the redshift interval $z=3-4$ and
$7.2\pm3.0\times10^{-7}$ counts~s$^{-1}$ for $z=4-5$.
The weighted mean $2-8$~keV counts~s$^{-1}$ in the combined fields
are weaker and only detected out to $z=3$, reflecting 
{\em Chandra}'s poorer sensitivity at these energies.

Converting the $0.5-2$~keV ($2-8$~keV) counts~s$^{-1}$
to flux with an average multiplication of 
$6.8\times10^{-12}$~erg~cm$^{-2}$ ($23.9\times10^{-12}$~erg~cm$^{-2}$),
where the conversion factors are taken from T11, we find the flux ratio 
$f(0.5-2$~keV)$/f(2-8$~keV)=$0.91\pm0.15$ at $z=0-1$, $0.95\pm0.23$ at $z=1-2$,
and $0.65\pm0.32$ at $z=2-3$. 
If the source spectra can be represented by power laws, then 
these flux ratios correspond to photon indices of $1.93\pm0.13$, 
$1.96\pm0.19$ and $1.69\pm0.28$, respectively, showing that 
the sources are quite soft on average.

\begin{inlinefigure}
\includegraphics[width=2.5in,angle=90]{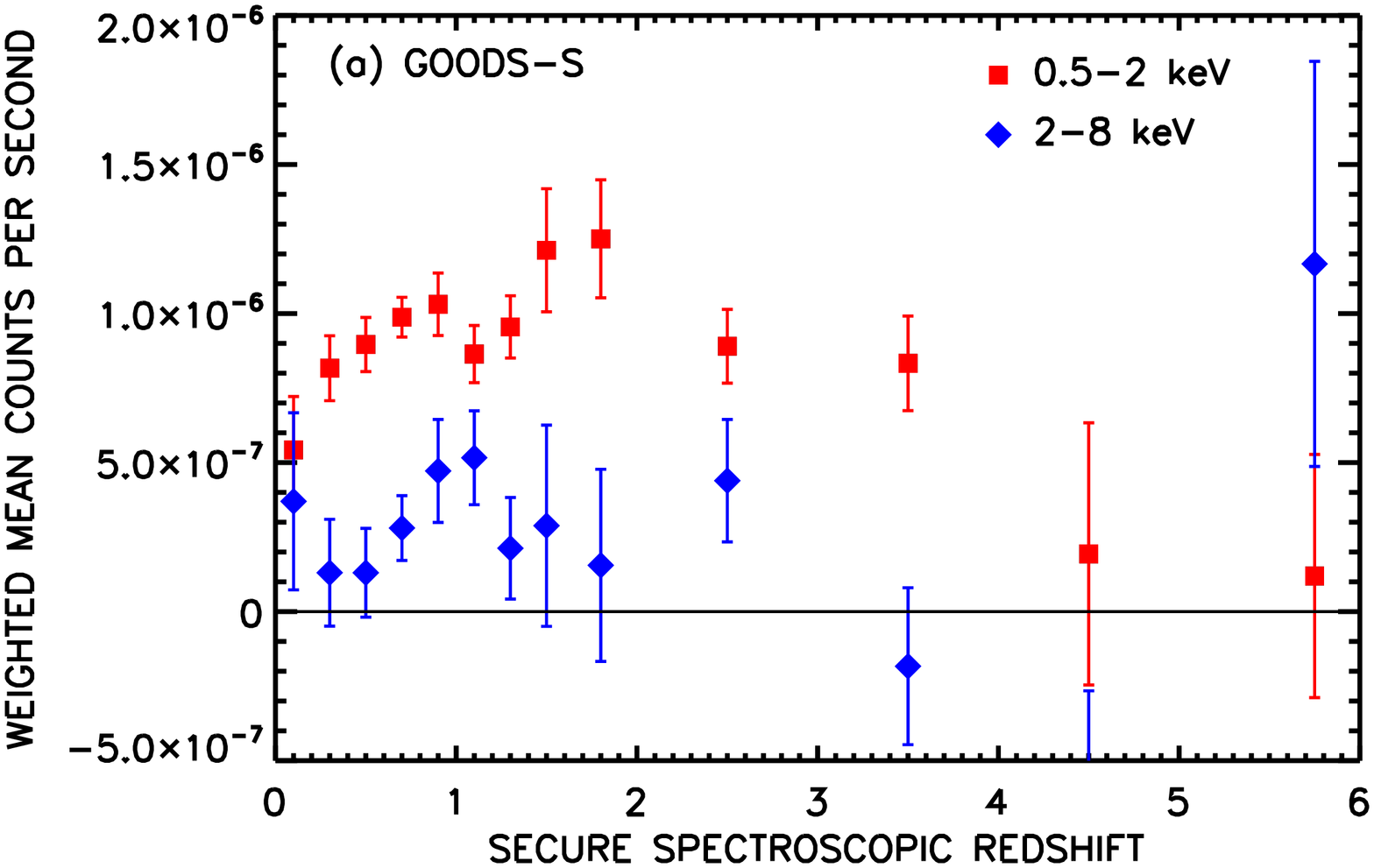}
\includegraphics[width=2.5in,angle=90]{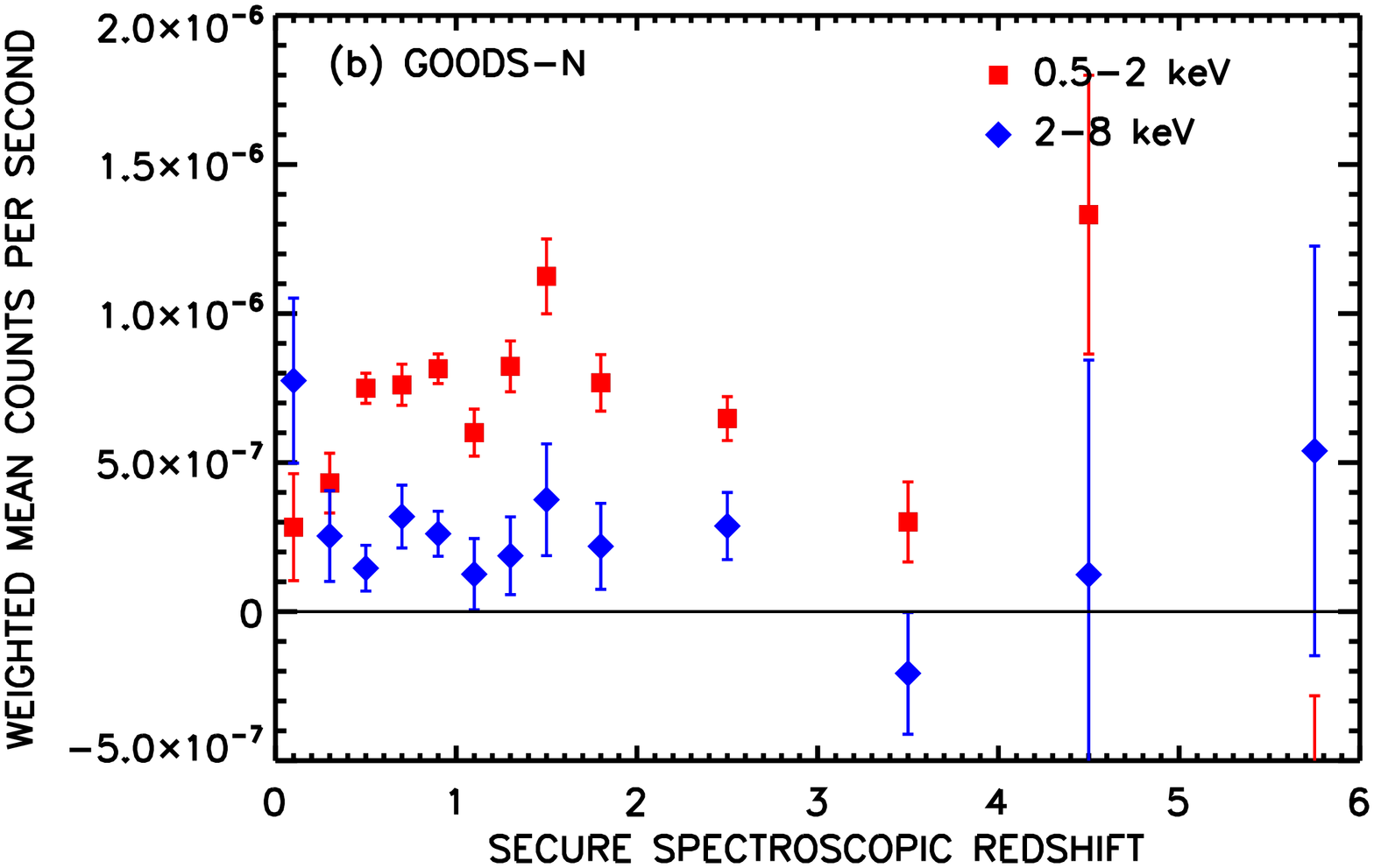}
\includegraphics[width=2.5in,angle=90]{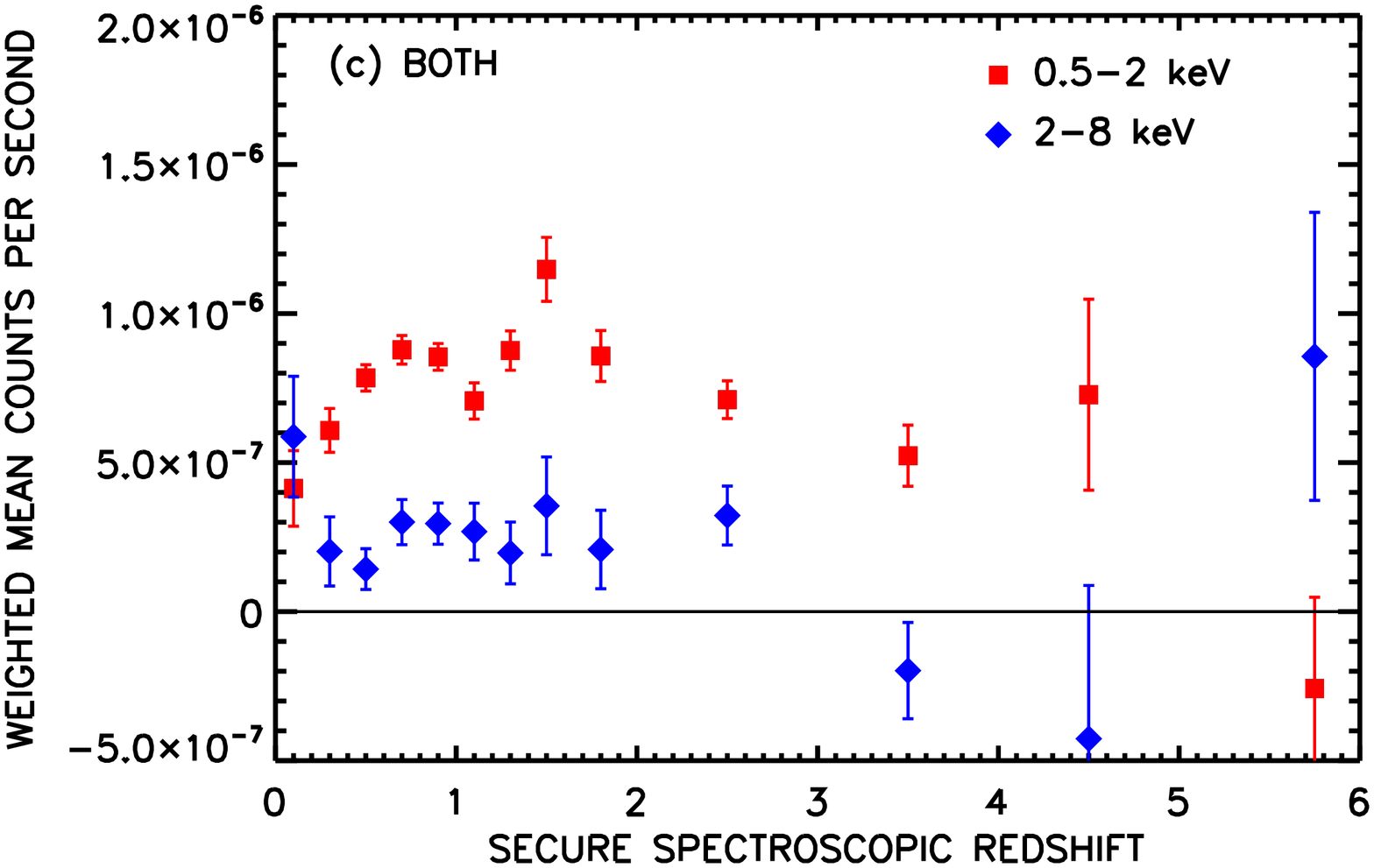}
\caption{
(a) 
Weighted mean X-ray counts~s$^{-1}$, $L$, vs. redshift for 
all sources with secure spectroscopic redshifts in 
(a) the GOOD-S core sample, (b) the GOODS-N core sample, 
and (c) the combined core samples.
The red squares denote the $0.5-2$~keV band,
and the blue diamonds denote the $2-8$~keV band.
Error bars are $\pm1\sigma$.
Sources directly detected in the X11 or the A03 X-ray catalogs 
are excluded.
\label{xray_redshift}
}
\end{inlinefigure}

\subsection{Photometric Redshift Sample}
\label{photzsamp}

In Figure~\ref{photz_redshift} we show the weighted mean 
counts~s$^{-1}$ in both the $0.5-2$~keV (red squares)
and $2-8$~keV (blue diamonds) bands versus redshift 
for the sources in the GOODS-S core region with photometric 
redshifts. We give the results in tabular form in Table~\ref{tab1},
where we compare with the GOODS-S secure spectroscopic redshift sample 
in the same redshift intervals. In Columns~(1) and (2), we give, 
respectively, the minimum and maximum of the redshift interval. 
In Column~(3), we give the number of sources in the spectroscopic 
redshift sample.  In Columns~(4) and (5), we give, respectively,
the mean $0.5-2$~keV and $2-8$~keV counts with $1\sigma$ errors
for the spectroscopic redshift sample in units of 
$10^{-7}$~counts~s$^{-1}$. In Column~(6), we give the number of 
sources in the photometric redshift sample.  Finally, in Columns~(7) 
and (8), we give, respectively, the mean $0.5-2$~keV and $2-8$~keV counts 
with $1\sigma$ errors for the photometric redshift sample in units of
$10^{-7}$~counts~s$^{-1}$.

\begin{deluxetable}{cccccccc}
\renewcommand\baselinestretch{1.0}
\tablewidth{0pt}
\tablecaption{Mean X-ray Counts Per Second Versus Redshift: GOODS-S}
\tablehead{
& & \multicolumn{3}{c}{Spectroscopic Sample} & 
\multicolumn{3}{c}{Photometric Sample} \\
$z_{\rm min}$ & $z_{\rm max}$ & $N_{\rm spec}$ & $0.5-2~{\rm keV}$ & $2-8~{\rm keV}$ & $N_{\rm photz}$ & $0.5-2~{\rm keV}$ & $2-8~{\rm keV}$  \\ (1) & (2) & (3) & (4)  & (5) & (6) & (7) & (8)}
\startdata
    0.0  &     0.50  &       127  &   7.64$\pm$0.83  &   2.59$\pm$1.37  &
    1170  &   1.59$\pm$0.28  &   1.41$\pm$0.46  \cr
   0.50  &      1.0  &       350  &   9.81$\pm$0.50  &   2.55$\pm$0.81  &
    2459  &   2.29$\pm$0.19  &  0.09$\pm$0.32  \cr
    1.0  &      1.5  &       190  &   9.39$\pm$0.68  &   3.95$\pm$1.11  &
    1500  &   3.35$\pm$0.25  &   1.50$\pm$0.41  \cr
    1.5  &      2.0  &        25  &   12.2$\pm$1.81  &  0.09$\pm$2.95  &
     817  &   3.76$\pm$0.34  &   1.76$\pm$0.55  \cr
    2.0  &      2.5  &        38  &   9.66$\pm$1.68  &   4.67$\pm$2.78  &
     835  &   3.00$\pm$0.33  &   1.52$\pm$0.54  \cr
    2.5  &      3.0  &        32  &   8.00$\pm$1.82  &   4.04$\pm$3.04  &
     697  &   2.10$\pm$0.36  &  0.15$\pm$0.58  \cr
    3.0  &      3.5  &        24  &   7.82$\pm$2.09  &   1.19$\pm$3.48  &
     430  &   2.46$\pm$0.46  &  0.21$\pm$0.76  \cr
    3.5  &      4.0  &        14  &   9.00$\pm$2.43  &  -5.86$\pm$4.02  &
     253  &   1.90$\pm$0.61  &  -1.03$\pm$0.99  \cr
    4.0  &      4.5  &         6  &  -1.64$\pm$4.57  &  -8.19$\pm$7.65  &
     145  &   1.00$\pm$0.81  & 0.07$\pm$1.33  \cr
    4.5  &      5.0  &         0  &  \nodata  &  \nodata  &        89  &
 1.98$\pm$0.986  &  0.79$\pm$1.62  \cr
    5.0  &      5.5  &         0  &  \nodata  &  \nodata  &        46  &
0.50$\pm$1.33  &  .015$\pm$2.19  \cr
    5.5  &      6.0  &         5  &   1.19$\pm$4.07  &   11.6$\pm$6.79  &
      31  &   2.02$\pm$1.70  &  0.46$\pm$2.81  \cr
    6.0  &      6.5  &         0  &  \nodata  &  \nodata  &         6  &
 3.64$\pm$3.59  &  -3.94$\pm$5.83  \cr
\enddata
\label{tab1}
\end{deluxetable}

Figure~\ref{photz_redshift} is similar to Figure~\ref{xray_redshift}
in that we see a strong signal out to $z=4$ in the $0.5-2$~keV band
but no detections at higher redshifts.
For example, for the redshift interval $z=4-5$,
the weighted mean $0.5-2$~keV counts~s$^{-1}$
is $1.38\pm0.62 \times10^{-7}$~counts~s$^{-1}$ 
or $f(0.5-2$~keV)=$9.4\pm4.2 \times 10^{-19}$~erg~cm$^{-2}$~s$^{-1}$.
We also again see a weak $2-8$~keV signal with respect to the 
$0.5-2$~keV signal; it is only detected in the redshift interval 
$z=1-3$, where we find a flux ratio 
$f(0.5-2$~keV)/$f(2-8$~keV)$=0.67\pm0.15$
corresponding to a photon index of $\Gamma=1.7$.

\begin{inlinefigure}
\centerline{\includegraphics[width=2.5in,angle=90]{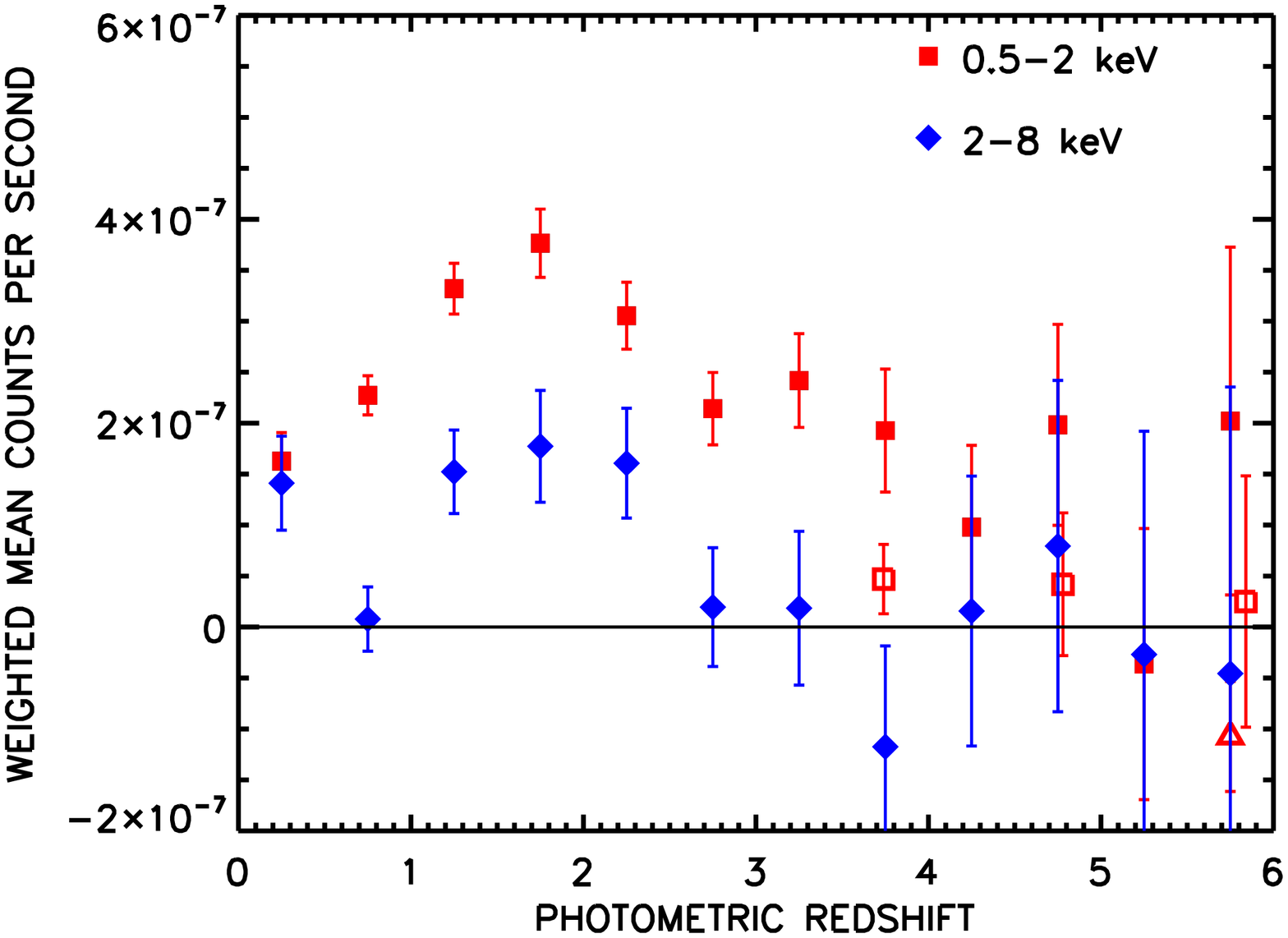}}
\caption{
Weighted mean X-ray counts~s$^{-1}$, $L$, vs. redshift for
the sources with photometric redshifts in the GOODS-S core region.
The red squares denote the $0.5-2$~keV band, and the blue diamonds 
denote the $2-8$~keV band. Error bars are $\pm1\sigma$.
Also shown are the weighted mean $0.5-2$~keV counts~s$^{-1}$ of 
the Beck06 $b$, $v$, and $i$ dropout samples
(red open squares; these samples are from the GOODS-S and HUDF) and 
the Bo06 $i$ dropout sample
(red open triangle; this sample is from the GOODS-N, GOODS-S, and HUDF). 
Sources directly detected in the X11 or the A03 X-ray catalogs
are excluded. None of the $z>4$ samples has any 
significant signal.
\label{photz_redshift}
}
\end{inlinefigure}

While the overall pattern with redshift is similar in the 
spectroscopic redshift and photometric redshift 
samples, the weighted mean X-ray counts~s$^{-1}$ are more 
than a factor of 2 lower in the photometric redshift sample 
than in the spectroscopic redshift sample. As we shall
show in the discussion section (Section~\ref{discuss}), 
this is a simple consequence
of the fact that the photometric redshift sample contains
sources down to fainter optical magnitudes than the spectroscopic
sample. The photometric
redshift sample, while containing the spectroscopically identified
sources as a subsample, probes to fainter optical magnitudes, and 
these optically fainter sources are intrinsically fainter in the 
X-rays, reducing the weighted mean X-ray fluxes.
However,
because the photometric redshift sample contains a larger 
number of sources, it identifies more of the X-ray background light, 
as measured by the product of the number of sources per unit area 
and the mean counts per source, than the secure spectroscopic
redshift sample. 
At redshifts between $z=0$ and 1, the contribution to
the X-ray background light by the photometric redshift
sample is about twice as much as the contribution to 
the X-ray background light by the secure spectroscopic sample, 
while between $z=2$ and 3, it is about six times as much.

The mean counts per source in the photometric redshift samples
do not change rapidly as a function of redshift.  Thus, the results
are not very sensitive to catastrophic photometric redshift
errors. If 4\% of the sources that the photometric redshift sample
placed at $z=0-1$ were really at $z=2-3$, then the mean counts at 
$z=2-3$ would only change from $2.59\times10^{-7}$~counts~s$^{-1}$ 
to $2.55\times10^{-7}$~counts~s$^{-1}$.
At $z=3-4$, the change in the mean counts would be even smaller.

\subsection{Dropout Samples}
\label{dropsamp}

None of the dropout samples are detected in either the
$0.5-2$~keV or the $2-8$~keV bands at any significant level. 
For the Beck06 $b$, $v$, and 
$i$ dropout samples, we find that the weighted mean
$0.5-2$~keV counts~s$^{-1}$ are
$4.7\pm3.4\times 10^{-8}$~counts~s$^{-1}$ for
$z=3.74$, $4.2\pm7.0\times 10^{-8}$~counts~s$^{-1}$ for $z=4.78$, 
and $2.5\pm12.3\times 10^{-8}$~counts~s$^{-1}$ for $z=5.84$.
We show these with red open squares on Figure~\ref{photz_redshift}.

For the Bo06 $i$ dropout sample, we find that the weighted mean
$0.5-2$~keV counts~s$^{-1}$ is 
$-10.7\pm5.3 \times 10^{-8}$ counts~s$^{-1}$
(red open triangle in Figure~\ref{photz_redshift}), 
while the weighted mean $2-8$~keV counts~s$^{-1}$
is $-13.1\pm9.7\times 10^{-8}$~counts~s$^{-1}$. 
In contrast, T11 obtained a $2-8$~keV signal
of $88\pm13\times 10^{-8}$~counts~s$^{-1}$, which is
highly inconsistent with the present limits. As we discussed
in Section~\ref{stack}, the T11 detection appears to be the
consequence of incorrect background subtraction.

At higher redshifts ($z\sim7-8$; Bo11 sample), we find that 
the weighted mean $0.5-2$~keV counts~s$^{-1}$ is 
$-11.4\pm5.1\times 10^{-8}$~counts~s$^{-1}$,
while the weighted mean $2-8$~keV counts~s$^{-1}$ is 
$-6.9\pm9.7\times 10^{-8}$~counts~s$^{-1}$.
We conclude that high-redshift sources are generally too
faint to be detected in X-rays, even in these extremely deep
stacks. The typical $2\sigma$ upper limits on the weighted mean 
fluxes are approximately $7\times10^{-19}$~erg~cm$^{-2}$~s$^{-1}$.

\section{Discussion}
\label{discuss}

The X-ray fluxes probed by the optimized averaging analysis
are, by construction, bounded
above by the flux detection limits in the direct X-ray catalogs.
As we illustrate in Figure~\ref{redshift_luminosity},
even at high redshifts, this places an upper bound
on the observed-frame $0.5-2$~keV
luminosities of $\sim 10^{42}$~erg~s$^{-1}$. We may therefore
expect the sources in the averaging analysis to be dominated by 
star-forming galaxies with some contribution from LLAGNs
(e.g., Bauer et al.\ 2004). The power law photon indices
based on the broadband X-ray colors ($\Gamma=1.7-2$
for $z=1-3$; see Section~\ref{specsamp}) are broadly consistent 
with this interpretation. Swartz et al.\ (2004) give an average 
$\Gamma=1.7$ for the ultraluminous X-ray 
sources (ULXs) that dominate the X-ray contributions in 
strong star-forming galaxies. However, the photon indices of 
ULXs are poorly determined above 10~keV, which adds some 
uncertainty to the comparison at the highest redshifts 
($z\sim3$) for which we were able to measure the weighted
mean observed-frame flux ratios, $f(0.5-2$~keV)$/f(2-8$~keV).
The dominance of star-forming galaxies at the low X-ray fluxes 
is also supported by fluctuation analyses, which show that the 
unresolved flux distribution is consistent with an extrapolation 
of the star-forming galaxy log N/log S 
(Hickox \& Markevitch 2007; Soltan 2011).

\begin{inlinefigure}
\centerline{\includegraphics[width=2.8in,angle=90]{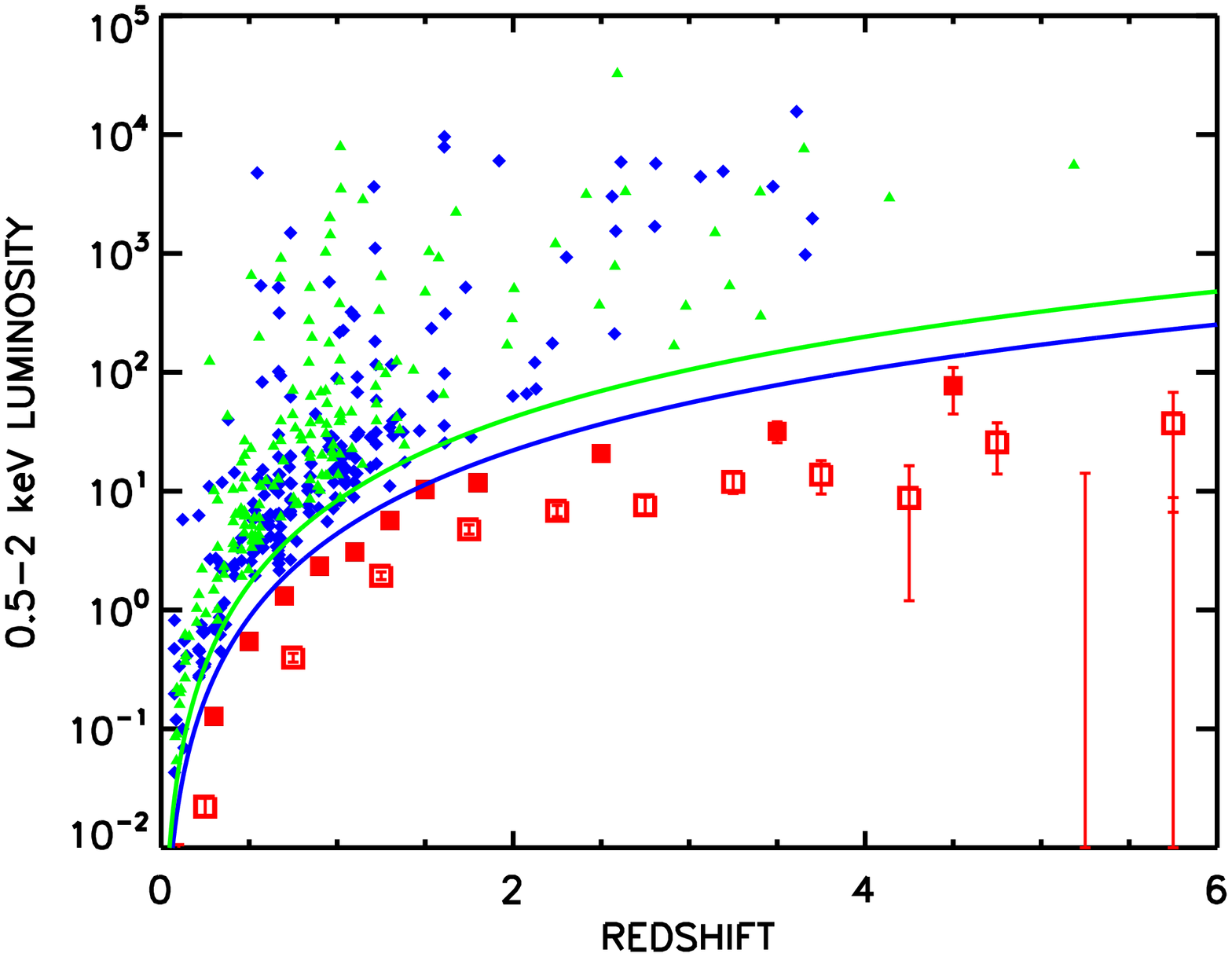}}
\caption{
Comparison of the weighted mean $0.5-2$~keV luminosities 
(units of $10^{40}$~erg~s$^{-1}$) of the sources in
the spectroscopic (red solid squares) and 
photometric (red open squares) redshift samples
with the individually detected sources in the CDF-S
(blue diamonds) and CDF-N (green triangles). 
Also shown are the on-axis detection limits for the CDF-S
(blue curve) and CDF-N (green curve).
\label{redshift_luminosity}
}
\end{inlinefigure}

Dijkstra et al.\ (2011) predicted the X-ray sky surface 
brightness (i.e., the X-ray background light) from
star-forming galaxies using the star formation history of
Hopkins \& Beacom (2006) and the star formation rate (SFR) to 
X-ray conversion of Mineo et al.\ (2011). 
This allows us to make a very simple comparison of the
light measured from the low-luminosity X-ray sources in the
averaging analysis with the X-ray prediction
made from star formation histories measured at other wavelengths.
In Figure~\ref{xray_surf_both} we compare the Dijkstra et al.\ 
calculation in the $2-8$~keV band (black solid curve; we assume
a photon index of $\Gamma=2$) with the 
X-ray contributions determined from the results of the present 
photometric redshift averaging analysis in the GOODS-S core region.
In Figure~\ref{xray_surf_both}(a) we use open squares to show
for the $0.5-2$~keV band the quantity 
$N~\times~S / (A \times (z_1-z_2))$, where $N$ is the 
number of sources used in the averaging analysis, 
$S$ is the weighted mean 
flux, $A$ is the observed area in square degrees, and 
($z_1-z_2$) is the redshift interval. We can combine this
with the contributions from the directly detected low-luminosity
sources in the X11 catalog to determine
the value for all sources with $0.5-2$~keV luminosities
less than $10^{42}$~erg~s$^{-1}$ (solid circles). 
Figure~\ref{xray_surf_both}(b) shows the same plot for the 
$2-8$~keV band.

\begin{inlinefigure}
\centerline{\includegraphics[width=2.8in,angle=90]{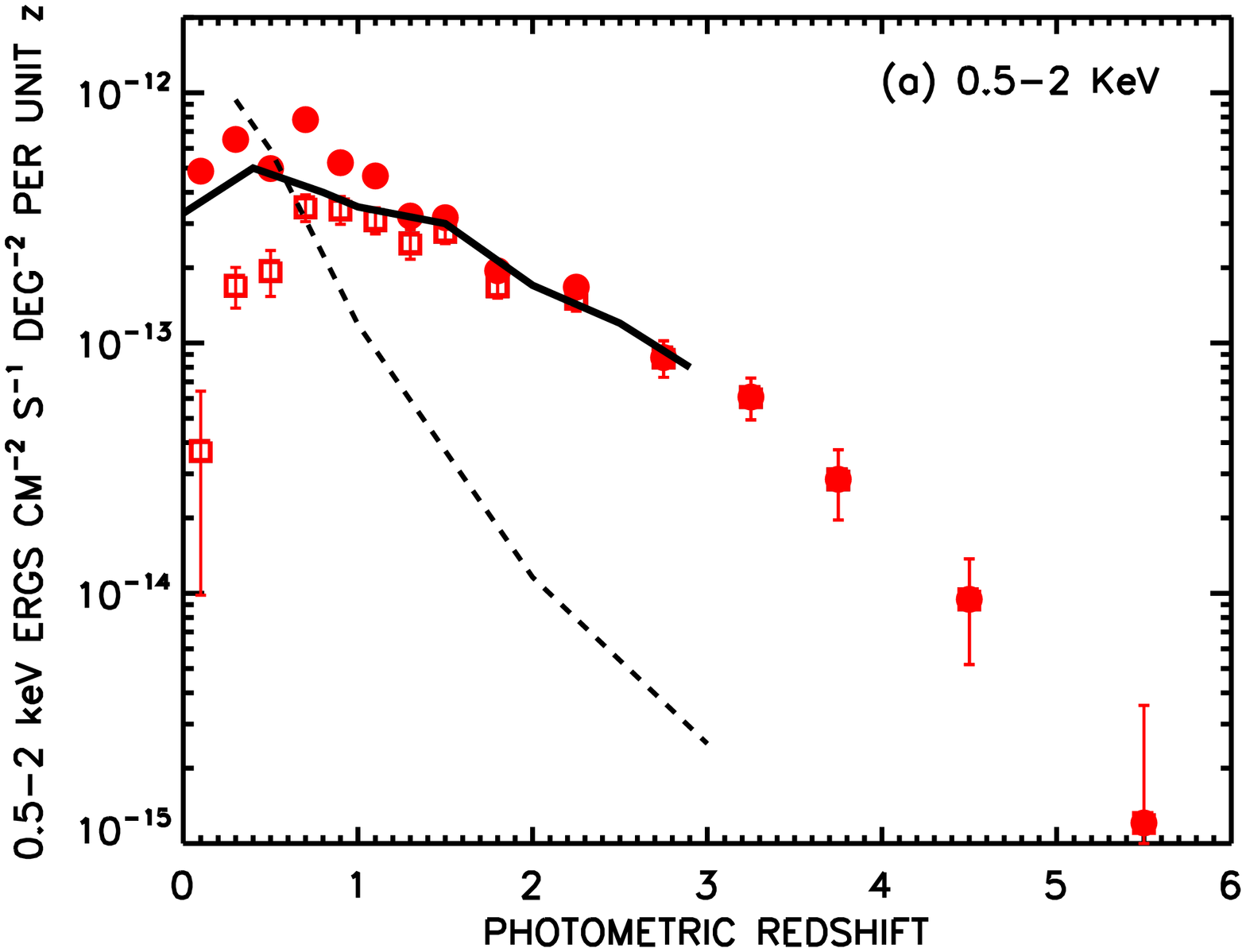}}
\centerline{\includegraphics[width=2.8in,angle=90]{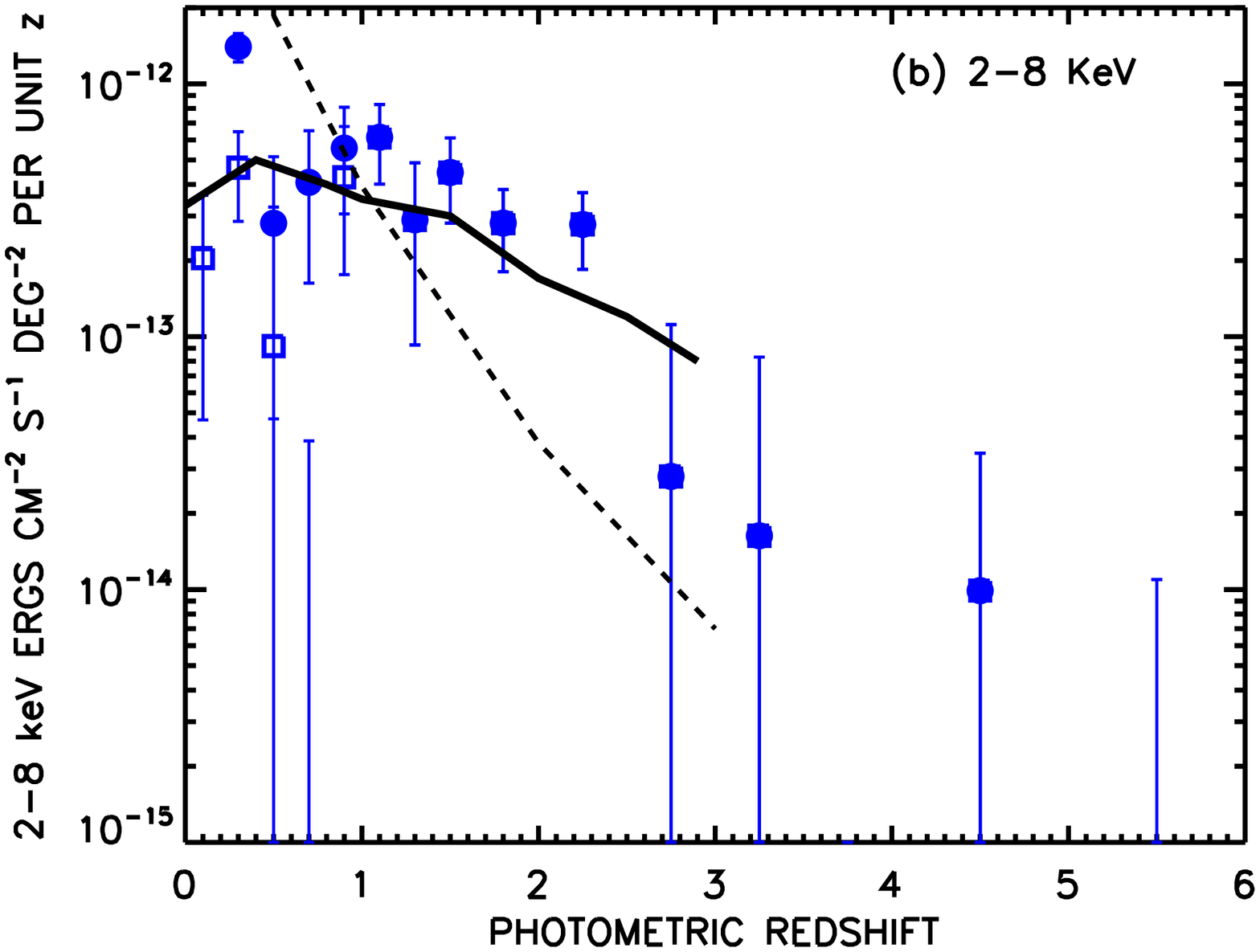}}
\caption{
(a) The $0.5-2$~keV contributions to the X-ray background determined 
from the results of the present photometric redshift averaging 
analysis in the GOODS-S core region (open squares).  These
contributions combined with the contributions from the directly 
detected low X-ray luminosity sources in the X11 catalog
give the total contributions to 
the $0.5-2$~keV surface brightness from sources with X-ray luminosities 
less than $10^{42}$~erg~s$^{-1}$ in the $0.5-2$~keV band (solid circles).
(b) Same for the $2-8$~keV contributions.
In both cases, the solid curve shows the Dijkstra et al.\ (2011) 
calculation in the $2-8$~keV band of the contribution of star-forming 
galaxies to the X-ray background 
(for the assumption of a photon index of $\Gamma=2$,
this should equal the contribution in the $0.5-2$~keV band), 
and the dashed curve shows the contribution of LLAGNs with X-ray 
luminosities less than $10^{42}$~erg~s$^{-1}$, as computed using the 
model of Gilli et al.\ (2007).
\label{xray_surf_both}
}
\end{inlinefigure}

The agreement between the averaging analysis results 
(open squares) and the Dijkstra et al.\ (2011) prediction
are extremely good. As expected from the discussion of the photon
indices, the $0.5-2$~keV and $2-8$~keV surface brightnesses
are almost identical and match to the prediction for most of 
the $z=0-3$ range over which the prediction was made,
though the noise levels are much higher in the $2-8$~keV sample.
The only exception is at low redshifts ($z<0.5$), where the upper 
bound on the flux excludes many of the star-forming galaxies. 
When the results from the directly detected low-luminosity 
sources in the X11 catalog are combined with the averaging 
analysis results (solid circles), the data also then 
agree with the prediction at these low redshifts.  
Integrating through the combined results,
we find that a very large percentage of the contribution to the
X-ray background from
$L_X<10^{42}$~erg~s$^{-1}$ star-forming galaxies
comes from low-redshift objects:
79\% at $0.5-2$~keV and 83\% at $2-8$~keV arise
from $z\lesssim2$. We also show the contributions
from LLAGNs with $L_X<10^{42}$~erg~s$^{-1}$ to
the $0.5-2$~keV and $2-8$~keV surface brightnesses,
as computed using the model of Gilli et al.\ (2007). These
can be significant at $z<1$ but drop rapidly at higher redshifts.

We may invert this and use the X-ray data to construct the star formation
history as a function of redshift from X-ray samples
(e.g., Norman et al.\ 2004; Lehmer et al.\ 2008).
We converted the total X-ray surface brightness 
in the $0.5-2$~keV band of all sources with
luminosities less than $10^{42}$~erg~s$^{-1}$ in each
redshift interval to the corresponding
comoving X-ray luminosity density in the $0.5-8$~keV
band, assuming a photon index of $\Gamma=2$.
We then computed the SFR per unit comoving 
volume using the relation in Mineo et al.\ (2011).
For the present data, we cannot separate the various contributions 
to the X-ray light, so rather than adopting the Mineo et al.\ (2011) 
best-fitting linear relation obtained from only the resolved 
sample (their Equation~22), we used their linear
relation obtained from only the unresolved galaxies (given in their
Section~8.1):
\begin{equation}
{\rm SFR(M}_{\odot}\ {\rm yr}^{-1}) = 2.7\times10^{-40} L_{0.5-8~{\rm keV}} 
({\rm erg\ s}^{-1}) \,.
\label{ran}
\end{equation}
Here the SFR is for a Salpeter (1955) initial mass
function stretching from 0.1 to 100~M$_\odot$. 
The normalization is a factor of 1.4 higher than if we had 
instead used their result for only the hard X-ray binaries.
We note that this calibration does depend on the star formation 
history.  We also note that there have been numerous calibrations of
the SFR with X-ray luminosity, and these vary by up to 40\%
(Grimm et al.\ 2003; Ranalli et al.\ 2003; 
Persic \& Rephaeli 2007; Lehmer et al.\ 2010; Mineo et al.\ 2011).  
Thus, the estimated SFR will be uncertain at least at this level. 

We show the star formation history
determined from the X-ray data in Figure~\ref{sfr_hist},
where we compare it with that derived by Hopkins \&
Beacom (2006) using multiwavelength data. The shape and normalization
of the two curves are in very good agreement. The normalization
agreement is probably overly good, since this could be changed
by any one of a number of factors, such as 
contamination by LLAGNs, cosmic variance, or the uncertainty
in the X-ray to SFR conversion factor.
As discussed above, the LLAGN contribution is probably
most significant below $z=1$, where it may result in the
X-ray estimate of the SFR being high.

\begin{inlinefigure}
\centerline{\includegraphics[width=2.8in,angle=90]{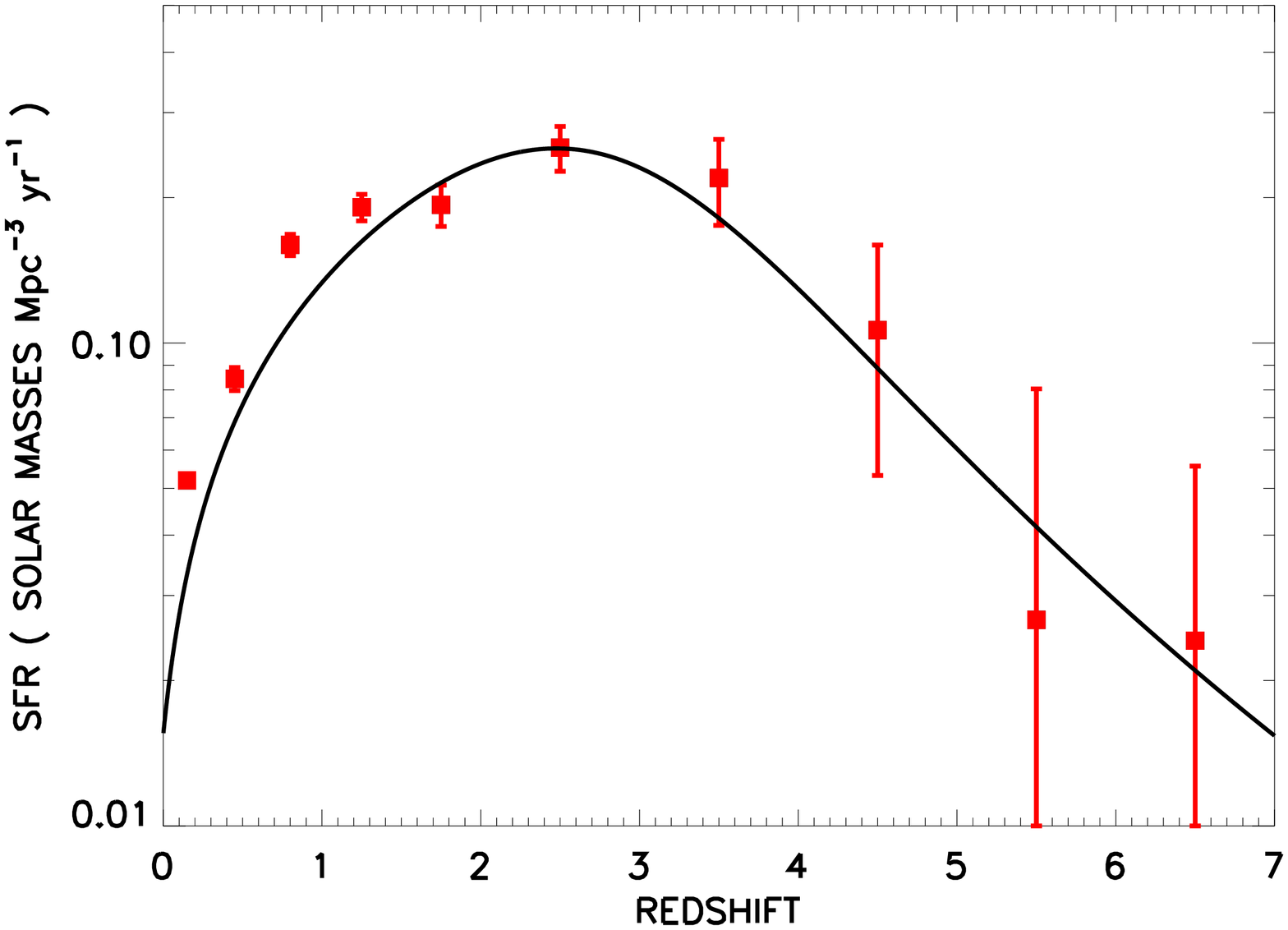}}
\caption{
The SFR per unit comoving volume calculated from
the $0.5-2$~keV sources with X-ray luminosities
less than $10^{42}$~erg~s$^{-1}$
(red squares with $1\sigma$ error bars) vs. redshift.
The black solid curve shows the Hopkins \& Beacom
(2006) parametric fit to their star formation history
renormalized to a Salpeter IMF stretching from
0.1 to 100~M$_\odot$.
\label{sfr_hist}
}
\end{inlinefigure}

With these data we can also address the evolution of the relation
between the X-ray luminosity and the SFR
as a function of redshift. This relation,
$L_X = c_X\times$ SFR, depends on the initial mass function
(IMF), the properties of binary stars, and the galaxy
metallicities, among other things. One might expect
that the normalization, $c_X$, would be higher in low-metallicity 
galaxies (e.g., Bookbinder et al.\ 1980;
Dray 2006; Linden et al.\ 2010). There is some weak evidence
for this (Kaaret et al.\ 2011). In turn, this might
suggest that $c_X$ could be higher at high redshifts.
Mirabel et al.\ (2010) invoke this evolution as a possible
reionization mechanism.

Dijkstra et al.\ (2011) tried to constrain such an evolution
using the contributions to the X-ray backgrounds.
Their results could only weakly constrain
any evolution in $c_X$, but this might be due to the fact that
the contributions to the backgrounds from the higher
redshifts are small.
We can more directly constrain any evolution
in $c_X$ as a function of redshift  by comparing how the X-ray flux
depends on the UV flux as a function of redshift. Since the
star formation histories, particularly at the highest
redshifts, are generally computed from extinction corrected
UV luminosity densities, this is the most direct test we
can make.

For each galaxy in the GOODS-S core region
with a photometric redshift,
we computed the rest-frame UV magnitude at 2500~\AA\
by interpolating between the four ACS bands in the 
GOOD-S catalog and then translating this to a bolometric flux,
$\nu f_\nu$, evaluated at the redshifted
wavelength. The rest-frame bolometric luminosity
at 2500~\AA\ is $\nu L_\nu = 4 \pi d_l^{2} \nu f_\nu$,
where $d_l$ is the luminosity distance.
We restricted our analysis to $z>1$
so that in our lower redshift intervals the
above interpolation is valid. At $z > 2.6$ where
the observed wavelength lies beyond the reddest
band (F850LP) in the ACS data, we used the
F850LP magnitude alone to estimate the rest-frame bolometric luminosity.

\begin{inlinefigure}
\includegraphics[width=2.5in,angle=90]{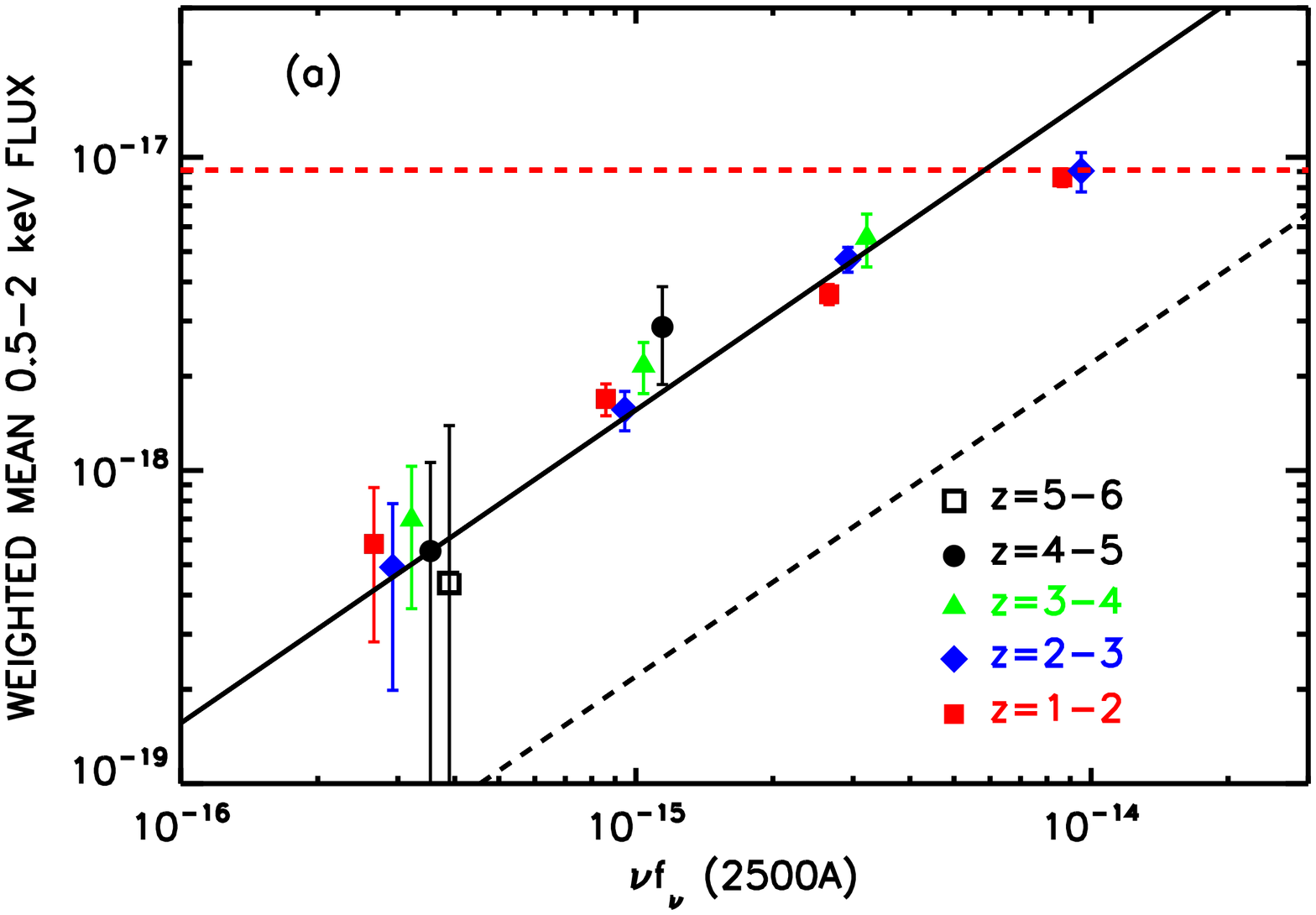}
\hskip -0.5cm
\includegraphics[width=2.5in,angle=90]{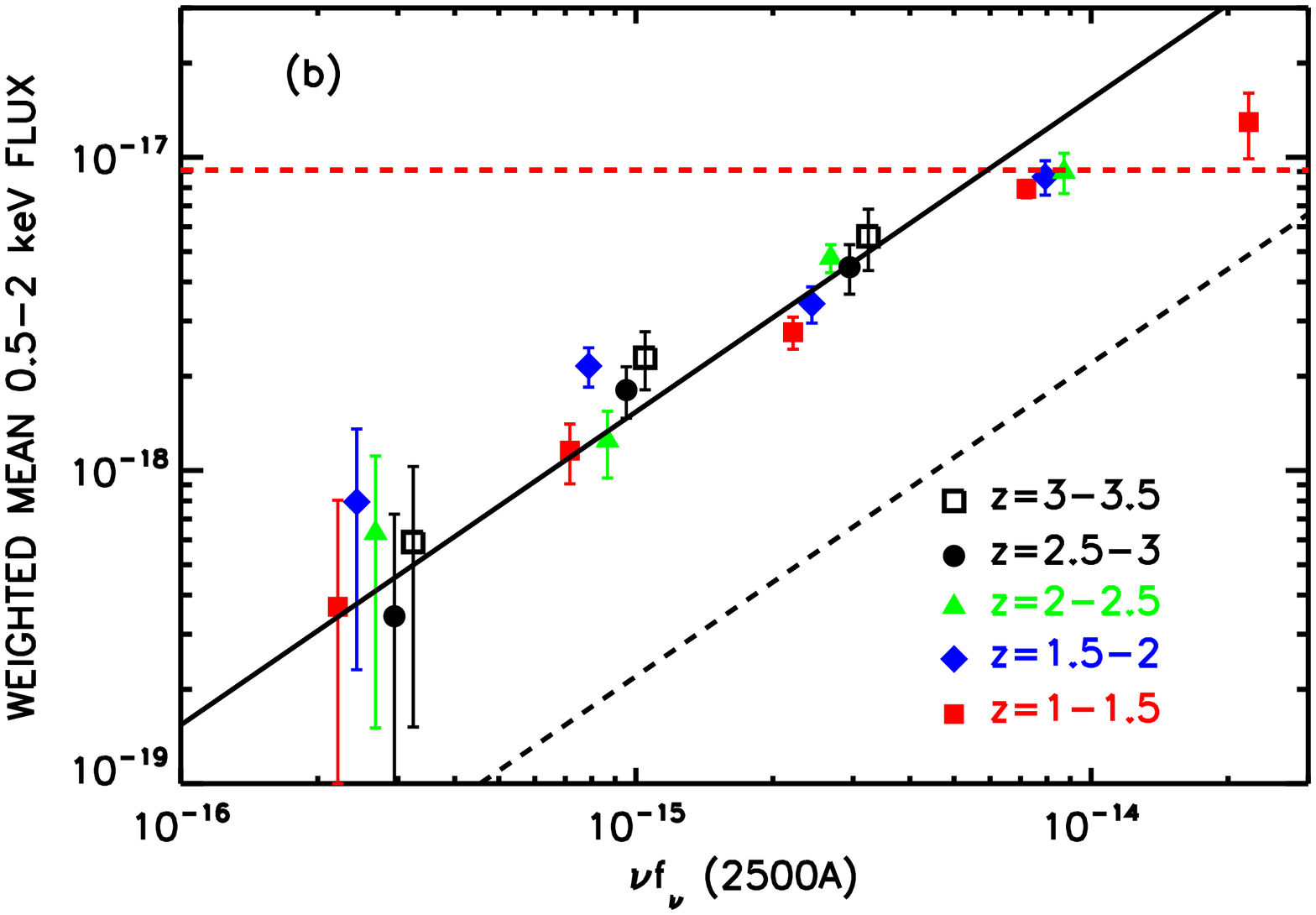}
\includegraphics[width=2.5in,angle=90]{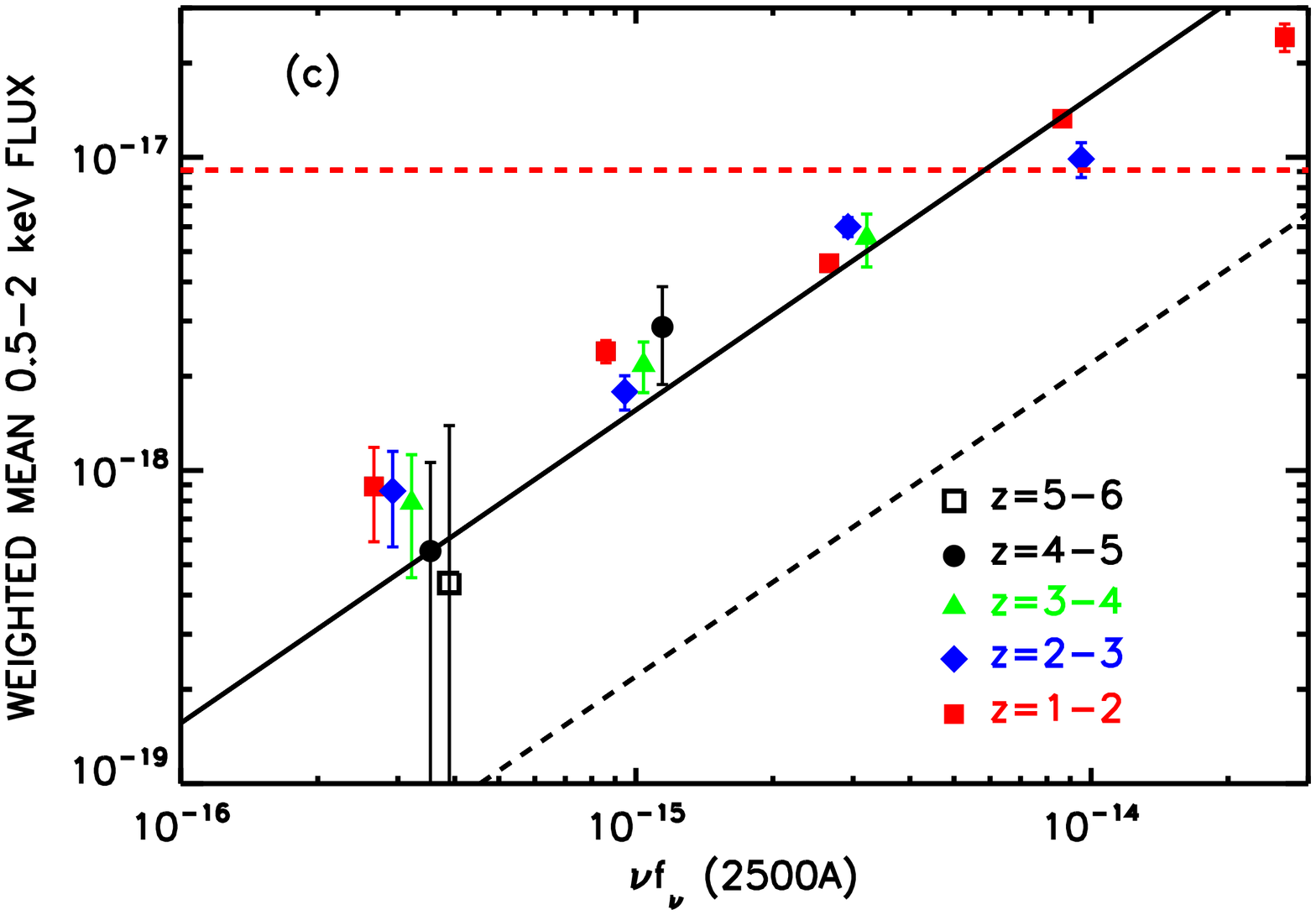}
\caption{
The weighted mean observed-frame $0.5-2$~keV flux of
the sources in the GOODS-S core region with photometric
redshifts vs. the bolometric flux,
$\nu f_{\nu}$, evaluated at a rest-frame wavelength
of 2500~\AA. If the X-ray source spectra are described by
power laws with $\Gamma=2$, then both axes
may simply be converted to luminosities by multiplying
by $4\pi d_l^2$.
(a) Correlation over $z=1-6$ for intervals of $dz=1$. 
(b) Correlation over $z=1-3.5$ for a finer gridding.
(c) Same as (a), but retaining in the averaging procedure
the directly detected low X-ray luminosity
($L_{0.5-2~{\rm keV}}<10^{42}$~erg~s$^{-1}$) sources in
the X11 catalog.
In all panels the flux ranges have been slightly
offset between redshift intervals for clarity.
The red dashed line shows the on-axis flux limit for a direct 
detection in the $0.5-2$~keV band. The black solid line shows
the linear relation 
$f_{0.5-2~{\rm keV}}=1.57\times10^{-3}~f_{\nu}\nu$(2500~\AA)
given in Equation~\ref{rel}.
The black dashed line shows the relation that would be obtained
using the local calibrations of Mineo et al.\ (2011)
for the X-ray luminosity vs. SFR and of Kennicutt (1998) 
for the near-UV luminosity vs. SFR, assuming 
$\Gamma=2$ and omitting any extinction correction.
\label{xray_uv}
}
\end{inlinefigure}

Using our optimal averaging procedure, we formed the weighted mean
X-ray fluxes in unit redshift intervals from $z=1-6$ and in $\nu f_\nu$
intervals stretching from $10^{-16}$~erg~cm$^{-2}$~s$^{-1}$
upwards in half dex intervals. We show the results
in Figure~\ref{xray_uv}, where we plot the weighted mean
observed-frame $0.5-2$~keV
flux against $\nu f_\nu$. For $\Gamma = 2$ there is no 
$K$-correction, and the X-ray luminosity in either the 
rest-frame $2-8$~keV band or the rest-frame $0.5-2$~keV 
band can be obtained from $4\pi d_l^{2} f_{0.5-2~{\rm keV}}$. 
Thus, in any redshift interval the plots may be considered as 
luminosity-luminosity plots.

Figure~\ref{xray_uv} illustrates a number of interesting points 
with regards to the X-ray fluxes. First, at any given redshift, there
is a linear relation between the weighted mean X-ray flux 
and $\nu f_\nu (2500$~\AA).
The linear relation only holds until we reach the flux level where we 
begin to exclude sources that are in the X11 catalogs
(red dashed horizontal line). This only affects
the high flux bins ($\nu f_\nu (2500$~\AA)$>
5\times10^{-15}$~erg~cm$^{-2}$~s$^{-1}$).
Retaining in the averaging procedure the directly detected
low X-ray luminosity 
($L_{0.5-2~{\rm keV}}<10^{42}$~erg~s$^{-1}$) sources in the X11
catalog, as in Figure~\ref{xray_uv}(c), extends
the relation for the low-redshift sources to higher fluxes.

The underlying reason for the linear relation between
the X-ray flux and $\nu f_\nu (2500$~\AA) is the linear relation 
between the X-ray flux and the SFR. However, since the conversion 
from the UV flux to a total SFR requires an extinction correction,
it also shows that the average extinction correction is similar for
all sources regardless of the near-UV flux. 

Second, there is little variation in the normalization with
redshift. For the full data set we find
\begin{equation}
f_{0.5-2~{\rm keV}} = 1.57\pm0.07\times10^{-3} \nu f_\nu (2500~{\rm \AA})
\label{rel}
\end{equation}
(black solid line).
However, fits to individual redshift intervals give almost
identical ratios ($1.48\pm0.09\times10^{-3}$ at $z=1-2$, 
$1.62\pm0.12\times10^{-3}$ at $z=2-3$, $1.89\pm0.24\times10^{-3}$ at $z=3-4$,
and $2.25\pm0.75\times10^{-3}$ at $z=4-5$), suggesting that any
increase with increasing redshift is small. More specifically,
the ratio is proportional to $\epsilon(z) c_X(z) (1+z)^{2-\Gamma}$,
where $\epsilon(z)$ is the mean extinction correction for the UV
flux and the $(1+z)^{2-\Gamma}$ factor arises from the $K-$correction 
for a power law spectrum with photon index $\Gamma$. For 
$\Gamma=2$ and $\epsilon(z)$ constant, the rise in $c_X (z)$ from 
$z=1.5$ to $z=4.5$ would be $1.5\pm0.5$, and for $\Gamma = 1.7$
and $\epsilon(z)$ constant, it would be $1.1\pm0.4$.

We may also compare the measured relation between $f_{0.5-2~{\rm keV}}$
and $\nu f_\nu (2500$~\AA) with that derived using local calibrations.
For this derivation we use the Mineo et al.\ (2011) calibration of 
the SFR with X-ray luminosity given in Equation~\ref{ran},
assuming a spectral index $\Gamma=$2 to convert the
$0.5-8$~keV luminosity to $0.5-2$~keV luminosity:
\begin{equation}
{\rm SFR(M}_{\odot}\ {\rm yr}^{-1}) = 5.4\times10^{-40} L_{0.5-2~{\rm keV}} 
({\rm erg\ s}^{-1}) \,,
\label{ran2}
\end{equation}
and the Kennicutt (1998) calibration of the SFR with the near-UV luminosity 
\begin{equation}
{\rm SFR(M}_{\odot}\ {\rm yr}^{-1}) = 1.17\times10^{-43} 
\nu L_{\nu} (2500~{\rm \AA}) ({\rm erg\ s}^{-1}) \,.
\label{ken}
\end{equation}
These are all computed for the same Salpeter (1955) IMF. Since we are 
only interested in the ratio, we only require a consistent choice of IMF. 
We combine Equations~\ref{ran2} and \ref{ken} to obtain the
relation
\begin{equation}
L_{0.5-2~{\rm keV}} = 2.2\times10^{-4} \nu L_\nu (2500~{\rm \AA})\,,
\label{rel2}
\end{equation}
which translates to 
\begin{equation}
f_{0.5-2~{\rm keV}} = 2.2\times10^{-4} \nu f_\nu (2500~{\rm \AA}) \epsilon(z) (1+z)^{2-\Gamma}\,,
\label{rel_fin}
\end{equation}
when we allow for the extinction correction, $\epsilon(z)$, to convert
from the UV-derived SFR to a total SFR, and when we allow for the 
$K-$correction.
In Figure~\ref{xray_uv} we show the result (black dashed line) 
obtained from this relation for $\Gamma=2$ when we do not correct
for extinction (i.e., $\epsilon(z) = 1$).
However, to match the observed flux, we need to assume that
$\epsilon(z) \sim 5$, which is very similar to values
estimated for the UV extinction corrections at these
redshifts such as Erb et al. (2006) who obtained a mean
correction of 4.5 for LBGs at $z=2$.
For $\Gamma=1.7$, the required 
extinction correction is reduced to $\epsilon(z)\sim3$.

We can see from Figure~\ref{xray_uv} that we can only marginally 
detect the most luminous star-forming galaxies
at $z=5-6$ with the present data.  The maximum 
$\nu f_\nu (2500$~\AA)$=4\times10^{-16}$~erg~cm$^{-2}$~s$^{-1}$
corresponds to 
$f_{0.5-2~{\rm keV}}=6\times10^{-19}$~erg~cm$^{-2}$~s$^{-1}$. 
A substantial detection of the $z=5-6$ sample using an optimal
averaging analysis of the present sort would require an
increase of about a factor of 3 in exposure time.

At higher redshifts ($z=6-8$) the $2\sigma$ upper limit on
$\nu f_\nu (2500$~\AA) corresponds to
$f_{0.5-2~{\rm keV}}=7\times10^{-19}$~erg~cm$^{-2}$~s$^{-1}$.
This translates to an upper limit on the X-ray luminosity 
at $z=6.5$ of $4\times10^{41}$~erg~s$^{-1}$ in the rest-frame
$3.75-15$~keV band. The highest redshift spectroscopically 
confirmed source in either the CDF-S or the CDF-N is at $z=5.19$ 
(Barger et al.\ 2003), and there are no known sources at higher
redshifts with luminosities above the threshold luminosities 
corresponding to the flux limits of the X-ray catalogs 
(see Figure~\ref{redshift_luminosity}).

Our upper limit on the X-ray luminosity at $z=6.5$
of $4\times10^{41}$~erg~s$^{-1}$ is a factor of 20 lower than the
luminosity given in T11, which was based on a supposed detection
in the less sensitive $2-8$~keV band.
Our detection limit is
consistent with the X-ray fluxes in the high-redshift sources
being solely due to star formation, though contributions
from LLAGNs could also be present. 
AGNs with luminosities above $10^{42}$~erg~s$^{-1}$ thus appear
to be extremely rare at high redshifts.

\acknowledgements

We thank the referee for useful comments that helped us to improve 
the manuscript.
We thank Ezequiel Treister for numerous discussions and 
for providing helpful information about the T11 paper.
We would also like to thank Dominik Bomans for several
stimulating conversations.
We gratefully acknowledge support from NSF grant 
AST-0709356 (L.~L.~C.), the University of Wisconsin
Research Committee with funds granted by the Wisconsin Alumni
Research Foundation and the David and Lucile Packard Foundation
(A.~J.~B.).
We wish to recognize and acknowledge the very significant
cultural role and reverence that the summit of Mauna Kea has always
had within the indigenous Hawaiian community.  We are most
fortunate to have the opportunity to conduct observations from
this mountain.


\end{document}